\documentclass[12pt]{article}
\usepackage{a4wide}
\usepackage{latexsym}
\usepackage{amsmath}
\usepackage{amssymb}
\usepackage{amsfonts}
\usepackage{amscd}
\usepackage{slashed}
\usepackage{cite}
\usepackage{placeins}
\usepackage{tikz-cd}
\usepackage[export]{adjustbox}
\usepackage{euscript}
\usepackage{caption}
\usepackage{subcaption}

\usepackage[pdfpagelabels=false]{hyperref}
\hypersetup{
	colorlinks,
	linkcolor={blue!50!black},
	citecolor={blue!50!black},
	urlcolor={blue!50!black}
}

\usepackage{pslatex}
\usepackage{graphicx}
\usepackage[latin1,utf8]{inputenc}
\usepackage[T1]{fontenc}

\RequirePackage{mathtools}
\usepackage{mathdots}
\usepackage{bm}
\usepackage{nomencl}

\usepackage{dcpic,pictex}

\numberwithin{equation}{section}
\allowdisplaybreaks

\newcommand{\eps}{\varepsilon}
\newcommand{\df}{\mathrm{d}}

\newcommand{\nm}{n_-}
\newcommand{\np}{n_+}

\def \be  {\begin{equation}}
\def \ee  {\end{equation}}
\def \ba  {\begin{eqnarray}}
\def \ea  {\end{eqnarray}}
\def \baa {\begin{eqnarray*}}
\def \eaa {\end{eqnarray*}}
\def \lab #1 {\label{#1}}

\def\d{\hbox{{d}\kern-.20em\hbox{l}}}

\def \matrix #1 {\left(\begin{array}{cc} #1 \end{array}\right)}

\def\II{\hbox{{1}\kern-.25em\hbox{l}}}

\newcounter{MBQ}

\DeclareMathAlphabet\mathbfcal{OMS}{cmsy}{b}{n}

\makenomenclature

\makeatletter
\newcommand*{\centerfloat}{%
  \parindent \z@
  \leftskip \z@ \@plus 1fil \@minus \textwidth
  \rightskip\leftskip
  \parfillskip \z@skip}
\makeatother

\begin{document}

\thispagestyle{empty}
\renewcommand{\thefootnote}{\fnsymbol{footnote}}

\begin{flushright}
  TUM-HEP-1550/25\\
  %arXiv:yymmnnnnn [hep-ph]
  February 4, 2025
\end{flushright}

\vspace{1.5cm}

\begin{center}
{\Large\bf Renormalization of the Next-to-Leading-Power Soft Function\\[0.1cm] for the Drell-Yan Off-diagonal Channel}\\
  \vspace{1cm}
  {\large Martin Beneke$^\dagger$, Yao Ji$^{\dagger}\footnote{Current address}$, Erik S\"underhauf$^\dagger$, and  Xing~Wang$^{\dagger\ast}$\\
  \vspace{0.5cm}
      {$^\dagger$\small \em Physik Department,} 
      {\small \em TUM School of Natural Sciences,} \\[-0.1cm]
      {\small \em Technische Universit\"at M\"unchen, James Franck-Stra\ss e 1,} 
      {\small \em D - 85748 Garching, Germany}\\[0.5em]
      {$^\ast$\small \em School of Science and Engineering, The Chinese University of Hong Kong, Shenzhen 518172, China}
  } 
\end{center}

\vspace{2cm}

% abstract ---------------------------------------
\begin{abstract}\noindent
We renormalize the soft function entering the factorization and resummation of the $qg$ parton-scattering channel of the Drell-Yan process near the kinematic threshold $\hat{s}\to Q^2$ at next-to-leading power 
in the expansion around $z \equiv Q^2 / \hat{s} = 1$, and solve its renormalization-group equation. 
\end{abstract}
\vspace*{\fill}

% -----------------------------------------------------------------------------

\newpage
\thispagestyle{empty}
\tableofcontents
\thispagestyle{empty}
% -----------------------------------------------------------------------------
\newpage
\setcounter{page}{1}

\renewcommand{\thefootnote}{\arabic{footnote}}
\setcounter{footnote}{0}

\section{Introduction}
\label{sect:intro}

In this paper we continue our investigation \cite{Beneke:2024cpq} of the renormalization of soft functions with soft field insertions and focus on a soft function that appears in the factorization of the Drell-Yan (DY) process near threshold $z\to 1$ at next-to-leading power (NLP) \cite{Beneke:2018gvs,Beneke:2019oqx,Broggio:2023pbu}. These soft functions are central elements for summing large logarithms at NLP. 
Specifically, we consider the soft function for the off-diagonal 
parton scattering $g+\bar{q}\to \gamma^*+X$ 
in the DY process, for which the factorization formula for the partonic cross section into a hard, two collinear and the soft function at NLP takes the form
\begin{equation}
	\label{eq:factnlpori}
	\Delta^{\rm NLP}_{g\bar{q}}(z)=2H(Q^2)\int\limits_{\rm sub}^\infty\frac{\df \omega_1}{\omega_1}\frac{\df\omega_2}{\omega_2} J_g(\omega_1)J^*_g(\omega_2)S^{\rm NLP}_{g\bar{q}}(\Omega; \omega_1, \omega_2),
\end{equation}
with $\Omega=Q(1-z)\ll Q$. The integration extends over positive $\omega_{1,2}$ with a subtraction at small $\omega$, which removes an endpoint divergence in the convolution \cite{Beneke:2019oqx,Broggio:2023pbu} and absorbs it into suitably defined threshold parton distributions. The details of the treatment of this endpoint divergence are not the subject of this paper, which is concerned with 
the evolution equation for the soft function $S^{\rm NLP}_{g\bar{q}}(\Omega; \omega_1, \omega_2)$.

The NLP soft functions with soft quark 
fields in addition to light-like Wilson lines are build from 
$\mathcal{P}_1(0,s\nm) = \left[0, s \nm\right]q_s(s\nm)$ for 
a colour-singlet interaction at space-time point $s\nm$, and 
\begin{equation}
\mathcal{P}_8^a(0,s\nm) = \left[0, s \nm\right]T^b\,q_s(s\nm)\,\mathcal{Y}^{ba}_{n_-}(s \nm)
\label{eq:octetblock}
\end{equation}
for the colour-octet case. Here $\left[0, s \nm\right]$ denotes a straight finite-distance light-like Wilson in the fundamental representation from the point $s\nm$ to 0, and  
$\mathcal{Y}^{ba}_{n_-}(s \nm)$ an adjoint Wilson line from  $-\infty$ to $s\nm$ along the light-like direction $\nm$.\footnote{Precise definitions are given in \eqref{eq:deltaadj} and \eqref{eq:Wilsonlength} below.}  
Multiplying $\mathcal{P}_1(0,s\nm)$ with the static anti-quark field 
$\bar{h}_v(0)$ results in the leading-twist heavy-quark light-cone distribution amplitude (LCDA), the anomalous dimension of which is well-known  \cite{Lange:2003ff,Braun:2003wx}. Multiplying two copies on different light rays, 
\begin{equation}
\mathcal{P}^\dagger_1(0,t \np)\frac{\slashed{n}_-\slashed{n}_+}{4}\mathcal{P}_1(0,s\nm)\,,
\label{eq:Higgs}
\end{equation}
gives the soft-quark function which appears in the factorization theorem for $\gamma\gamma\to h$ through light quark loops \cite{Liu:2019oav}. The anomalous dimension factorizes into the sum of two identical copies for the two light-cones and is related to all orders to the one of the leading-twist LCDA, after accounting for the local heavy-quark anomalous dimension \cite{Beneke:2024cpq}. The constraints from conformal symmetry can therefore be leveraged for the soft-quark function relevant to Higgs production. 

The colour-octet operator $\mathcal{P}_8^a(0,s\nm)$ arises when an energetic gluon emits a soft anti-quark and thereby converts to an energetic quark,  
which then undergoes a hard interaction at $x=0$. The incoming gluon from 
asymptotic $-\infty$ generates the semi-infinite adjoint Wilson line in 
\eqref{eq:octetblock}, which substantially changes the nature of the renormalization problem. The presence of semi-infinite Wilson lines induces 
IR / rapidity divergences, which must be regulated non-dimensionally 
in order to extract the ultraviolet behaviour that defines the anomalous dimension from the $1/\epsilon$ poles in dimensional regularization. However, the pole part is then found to depend on the IR regulator.
This complication is not specific to QCD, but appears generally when 
energetic charged particles enter from or move towards infinity. In the treatment of soft QED radiation from the leptons in the rare B-meson decay $B_s\to\ell^+\ell^-$ \cite{Beneke:2019slt} a subtraction of the soft operator was introduced to make the anomalous dimension well-defined, which induces a rearrangement of terms between the soft function of interest and collinear functions in the factorization formula for the process under consideration. The same method was then applied to soft photon radiation from energetic charged pions described by their LCDA \cite{Beneke:2020vnb}. 
For the non-abelian case, the analogous subtraction was shown \cite{Beneke:2024cpq} to also lead to a well-defined and gauge-invariant anomalous dimension for the soft function relevant to $gg\to h$, given by the 
generalization of \eqref{eq:Higgs} with $\mathcal{P}_1(0,s\nm)\to \mathcal{P}_8^a(0,s\nm)$.

The soft function for the off-diagonal DY production channel considered here captures the soft physics at the amplitude-squared level and requires a generalization of the subtraction procedure. We now deal with a soft 
operator of the form 
\begin{equation}
\mathcal{P}^{a\dagger}_8(x_0,x_0+t \nm)\frac{\slashed{n}_-}{4}\mathcal{P}^a_8(0,s\nm)\,.
\label{eq:NLPsoftoperatordefschematic}
\end{equation}
Unlike for $gg\to h$, the DY soft-quark function 
is the ``square'' of two building blocks on the same light-cone, separated by a time-like distance due to the inclusive sum over the soft final state. This soft function therefore depends on three variables $s$, $t$ and $x^0$. As shall be seen, 
unlike the vacuum matrix element of this operator, which has been computed to $\mathcal{O}(\alpha_s^2)$ \cite{Broggio:2023pbu}, the dependence of the anomalous dimension on these three variables turns out to be remarkably simple.

For the DY process the soft subtraction is in fact already relevant for the leading-power (LP) soft function, as emphasized in \cite{Chay:2013zya}. The issue has almost always been ignored in previous work starting from \cite{Korchemsky:1993uz}. While this can be justified and leads to the correct result at LP, this holds no longer true beyond. In Sec.~\ref{sect:softLP}, we therefore revisit the treatment of the LP DY soft function within the $\delta$ IR-regulator scheme for semi-infinite Wilson lines (following \cite{Beneke:2019slt}) in order to identify the necessary IR subtraction that can be applied at NLP as well. We then, in Sec.~\ref{sect:softNLP}, compute the anomalous dimension of the NLP DY soft-quark function, which is the first case of an amplitude-squared soft function beyond LP for which its renormalization is studied. The renormalization kernel turns out to have a completely factorized form. In Sec.~\ref{sect:endpoint} we make 
the observation that the convolution of the found soft-function kernel with 
collinear functions is expected to be endpoint-finite. Finally, in Sec.~\ref{sect:RGEsol} we provide the general solution to the 
renormalization-group equation through Mellin transformation and undertake a 
numerical study of the scale-evolution of the tree-level soft function. 
We conclude in Sec.~\ref{sect:conclusions}. Several appendices provide 
additional results and further technical details. We particularly highlight Appendices~\ref{appendix:subopandPDF} and~\ref{app:Bt3-to-hgg}.  The former puts the IR rearrangement for the soft function into the context of the factorization theorem and demonstrates that it is consistent with the standard renormalization of the parton distributions in the $\delta$ IR-regulator scheme. The latter establishes a relation between 
the kernel for the twist-3 B-meson LCDA in the soft limit and the kernels related to the structure \eqref{eq:octetblock}, which appear in 
the soft-quark function relevant to Higgs production and  
DY production.

%%%%%%%%%%%%%%%%%%%%%%%%%%%%%%%%%%%%%%%%%%%%%%%%%%%%%%%%%%%%%%%%%%%%%%%

\section{Leading-power soft function revisited}
\label{sect:softLP}

%%%%%%%%%%%%%%%%%%%%%%%%%%%%%%%%%%%%%%%%%%%%%%%%%%%%%%%%%%%%%%%%%%%%%%%
\begin{figure}[t]
\centering
\includegraphics[width=14cm]{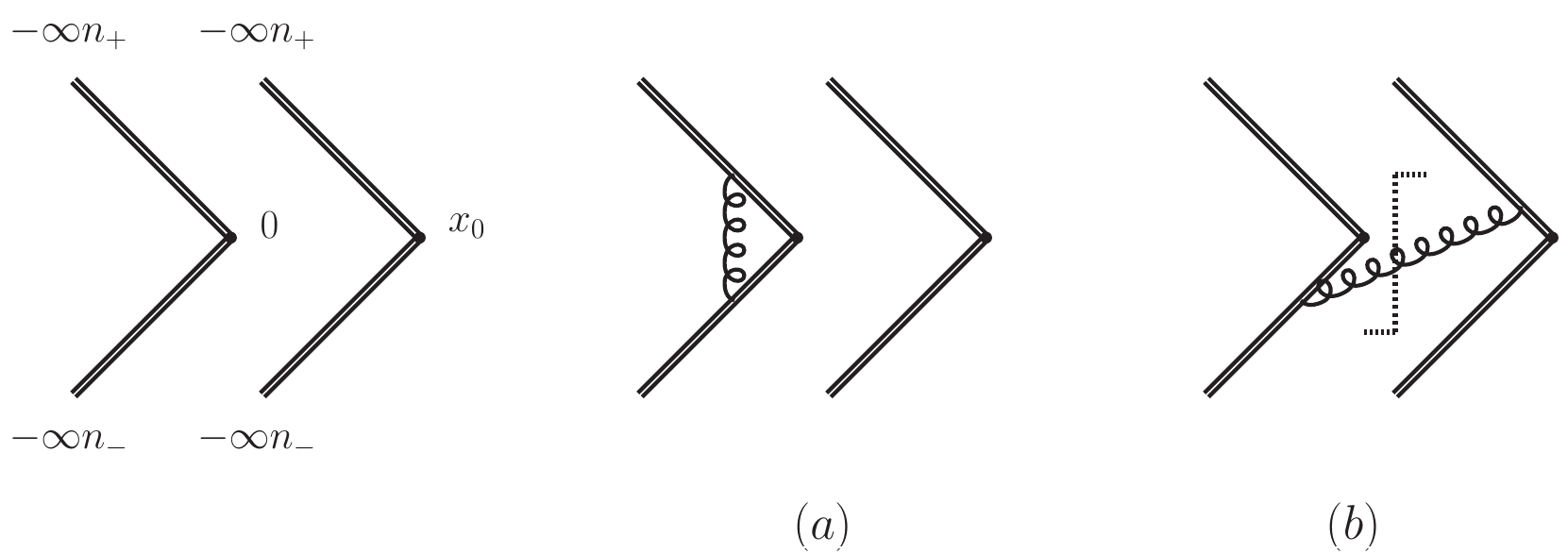}
\caption{\label{fig:LP}Leading-power Drell-Yan soft function and its one-loop corrections. Mirror diagrams of $(a)$ and $(b)$ are not displayed.}
\end{figure}
%%%%%%%%%%%%%%%%%%%%%%%%%%%%%%%%%%%%%%%%%%%%%%%%%%%%%%%%%%%%%%%%%%%%%%%

At LP the partonic cross section of the diagonal DY production channel $q\bar{q}\to \gamma^*$ factorizes near threshold $\hat{s}\to Q^2$, 
where $\hat{s} \equiv Q^2/z$ denotes the partonic centre-of-mass energy 
\begin{equation}
	\label{eq:factLP}
	\Delta^{\rm LP}_{q\bar{q}}(z)=2H(Q^2)S^{\rm LP}_{q\bar{q}}(\Omega),
\end{equation} 
with the same hard function as in \eqref{eq:factnlpori}. 
The soft function is defined as
\begin{equation}
\label{eq:LPsoftdef}
	S_{q\bar{q}}^{\rm LP}(x_0) =  \frac{1}{N_c}\,\langle0|\mbox{Tr} \, \mathbf{\bar{T}}(Y^\dagger_{n_-}(x_0) Y_{n_+}(x_0)) \,\mathbf{T}(Y^\dagger_{n_+}(0) Y_{n_-}(0))|0\rangle\,,
\end{equation}
with semi-infinite light-like Wilson lines
\begin{equation}
\label{eq:Wilsonlines}
	Y_{n_\pm}(x) = \widehat{\rm P}\exp\left[ig_s\,T^a\int_{-\infty}^{0 }\df 
\lambda\,n_\pm\cdot A^a(x+\lambda n_\pm)\right].
\end{equation}
Here $\textbf{T} (\mathbf{\bar{T}})$ denotes (anti-)time ordering, 
$\widehat{\rm P}$ path-ordering (turning into anti-path ordering in the hermitian conjugate). 
The vectors $n_\pm$ are two light-like vectors satisfying $\nm\cdot \np=2$,  whereas the position $x_0$ in \eqref{eq:LPsoftdef} is time-like, $x_0=(x^0,\vec{0})=\frac{x^0}{2}(\nm+\np)$. (We denote the vector as $x_0$ and its zero-component as $x^0$.) The analytic structure requires that $x^0/2$ is 
understood as $x^0/2-i0^+$, where $i0^+$ denotes an infinitesimal positive imaginary ``number''.

The operator and its one-loop corrections are shown in Fig.~\ref{fig:LP}. If dimensional continuation to $d=4-2\epsilon$ is employed to regulate ultraviolet (UV) and infrared (IR) singularities, the virtual correction $(a)$ is scaleless, while the real emission graph $(b)$ and its mirror are non-vanishing with poles in $\epsilon$ of IR origin. Due to the scaleless integrals in dimensional regularization, however, the $\epsilon$ poles in the sum of $(a)$ and $(b)$ are 
effectively UV poles. The one-loop result reads \cite{Korchemsky:1993uz}
\begin{eqnarray}
	\label{eq:LPsoft}
		S_{q\bar{q}}^{\rm LP}(x_0) 
  %\equiv \langle 0 |O_{q\bar{q}}^{\rm LP}(x_0) | 0 \rangle 
  &=& 1 + \frac{\alpha_s C_F}{\pi } \frac{e^{-\eps\gamma_E}\Gamma(1-\eps)}{\eps^2}\left(i\mu e^{\gamma_E}(x^0/2-i0^+)\right)^{2\eps}+\mathcal{O}(\alpha_s^2)\nonumber\\
		&=& 1 + \frac{\alpha_s C_F}{\pi }\left[ \frac{1}{\eps^2}+\frac{2L_{x_0}}{\eps}+2L_{x_0}^2+\frac{\pi^2}{12}+\mathcal{O}(\eps)\right]+\mathcal{O}(\alpha_s^2),
\end{eqnarray}
with
\begin{equation}
	\label{eq:Lx0}
	L_{x_0}=\ln\left(i\mu e^{\gamma_E}(x^0/2-i0^+)\right).
\end{equation}
Here and throughout, $\alpha_s$ denotes the $\overline{\rm MS}$ coupling at scale $\mu$. 
The $-i0^+$ prescription on $x^0/2-i0^+$ implies that the Fourier-transformed soft function\footnote{In the following, we will suppress the $-i0^+$ prescription for $x^0/2$ and $t$ appearing in \eqref{eq:Lx0} and below unless necessary. We will also use the same symbol to denote a  function and its Fourier-transform. The functions in position and  momentum space are distinguished by their arguments.}
\begin{equation}
	\label{eq:LPsoftmomdef}
	S_{q\bar{q}}^{\rm LP}(\Omega) = \int\limits_{-\infty}^{+\infty}\frac{\df x^0}{4\pi}e^{i\Omega x^0/2}S_{q\bar{q}}^{\rm LP}(x_0)
\end{equation}
has positive-energy support due to
\begin{equation}
	\label{eq:FourierOmega}
	\int\limits_{-\infty}^{+\infty}\frac{\df y}{2\pi}e^{i\Omega y}\big[i\left(y-i0^+\right)\big]^{\eta} = \frac{\Omega^{-1-\eta}}{\Gamma(-\eta)}\,\theta(\Omega)\quad\mbox{for}\quad -1 < \mbox{Re}(\eta) < 0\,.
\end{equation}

At next-to-leading power, however, one needs to distinguish $\eps$ poles of IR origin from those of UV origin for the purpose of renormalizing the soft operator. 
We therefore introduce a non-dimensional IR regulator to single out 
the UV poles in $\eps$ by introducing  $\delta$-regulators on the semi-infinite Wilson lines \cite{Beneke:2019slt,Beneke:2024cpq}. This regulator has the advantage that in the context of a factorization formula with soft and collinear functions, it is related to an off-shell regulator for the external energetic massless particles in the full theory and therefore descends consistently to the soft and collinear functions, which allows their computation with the method of regions \cite{Beneke:1997zp}, by-passing explicit matching. Explicitly, we define the regulated semi-infinite Wilson lines that arise from soft emissions of incoming particles in the fundamental representation as follows: 
\begin{equation}
\label{eq:delta}
Y_{n_\pm}(x) = \widehat{\rm P}\exp\left[ig_s\,T^a\int_{-\infty}^{0 }\df \lambda\,e^{-i\lambda (\delta_\pm+i0^+)}\,n_\pm\cdot A^a(x+\lambda n_\pm)\right],
\end{equation}
with $\delta_\pm$ real-valued. In the following, however, it will be convenient to include the $+i0^+$-prescription into the definition of 
$\delta_\pm$, and to introduce complex conjugation, i.e.~$\delta_\pm^*$, which should be understood as
\begin{equation}
    \label{eq:deltapmconjugate}
    \delta_-^* \longrightarrow \delta_- - i0^+,\quad \delta_+^* \longrightarrow \delta_+ - i0^+.
\end{equation}

For later purposes, we note the definition of the semi-infinite Wilson lines in the adjoint representation\footnote{We remark that here we adopt a different convention for the adjoint Wilson line in comparison to Eq.~(2.6) of Ref.~\cite{Beneke:2024cpq}, namely ${\cal Y}_{\text{Ref.~\cite{Beneke:2024cpq}}}= {\cal Y}^T{\text{this work}}$. We also note that in the presence of the $\delta$-regulator, the adjoint Wilson line is no longer strictly real, hence we distinguish $\mathcal{Y}^\dagger$ from $\mathcal{Y}^T$.}
\begin{equation}
\label{eq:deltaadj}
		\mathcal{Y}^{ab}_{n_\pm}(x) = \widehat{\mathrm{P}} \exp \left[g_s f^{cab} \int_{-\infty}^{0} \df\, \lambda \,e^{-i\lambda (\delta_\pm+i0^+)}n_\pm \cdot A^c\left(x+\lambda n_\pm\right)\right].
\end{equation}
As before, we will use $\delta_\pm$ and $\delta_\pm^*$ for brevity. 
The finite-distance Wilson line is obtained from two semi-infinite Wilson lines and reads 
\begin{equation}
	\label{eq:Wilsonlength}
	\left[u_1 n_{ \pm}, u_2 n_{ \pm}\right] = Y_{n_\pm}(u_1 n_\pm)Y_{n_\pm}^\dagger(u_2 n_\pm) = \widehat{\mathrm{P}} \exp \left[i g_s T^a \int_{u_2}^{u_1} \df\, \lambda\, n_{ \pm}\cdot A^a\left(\lambda n_{ \pm}\right)\right]\,.
\end{equation}
The same definition applies to the adjoint representation by substituting the colour generator $(T^a)_{bc} = -if^{abc}$.

\subsection{Unsubtracted soft function}
\label{subsect:LPunsub}

To validate the formalism and identify the soft subtraction for the amplitude-squared soft functions that appear in parton scattering, we first revisit the leading-power soft function in this context. We start with the unsubtracted, $\delta$-regulated LP DY soft function and then turn to the subtractions, which remove the $\delta$-regulator dependence. The unsubtracted soft function is gauge-dependent. The following results are obtained in Feynman gauge. We provide the  expressions in general $R_\xi$-gauge and discuss the gauge independence of the soft function after subtraction in App.~\ref{appendix:gauge}. 

With $\delta$-regulated semi-infinite Wilson lines the virtual correction $(a)$ in Fig.~\ref{fig:LP} is no longer scaleless, instead 
\begin{equation}
	\label{eq:LPsoftunsapre}
	\begin{aligned}
		S_{q\bar{q},{\rm uns}}^{{\rm LP}, (a)}(x_0) &= -\frac{\alpha_s C_F}{2\pi}e^{\eps\gamma_E}\Gamma(1-\eps)\Gamma^2(\eps)\left(\frac{\mu}{\delta_-}\right)^{\!\eps}\left(\frac{\mu}{\delta_+^*}\right)^{\!\eps}.
	\end{aligned}
\end{equation}
The mirror contribution of $(a)$ is the complex conjugate, hence simply replace $\delta_+\leftrightarrow\delta_-$. As a result, the virtual correction is real-valued and given by 
\begin{align}
	\label{eq:LPsoftunsv}
	S_{q\bar{q},{\rm uns}}^{{\rm LP}, \,v}(x_0) &= -\frac{\alpha_s C_F}{2\pi}e^{\eps\gamma_E}\Gamma(1-\eps)\Gamma^2(\eps)\left[\left(\frac{\mu}{\delta_-}\right)^{\!\eps}\left(\frac{\mu}{\delta_+^*}\right)^{\!\eps}+\left(\frac{\mu}{\delta_-^*}\right)^{\!\eps}\left(\frac{\mu}{\delta_+}\right)^{\!\eps}\right]\\
    &= -\frac{\alpha_sC_F}{\pi}\left[\frac{1}{\eps^2}+\frac{1}{\eps}\left(\ln\frac{\mu}{|\delta_-|}+\ln\frac{\mu}{|\delta_+|}\right)+\mathcal{O}(\eps^0)\right].
    \label{eq:LPsoftunsv2}
\end{align}

The real emission diagrams do not have UV poles in $\eps$. Hence, if one is only interested in the anomalous dimension, they can be discarded. However, for later discussion we provide them here:
\begin{eqnarray}
		S_{q\bar{q},{\rm uns}}^{{\rm LP}, (b)}(x_0) &=& 4\pi g_s^2C_F\int\limits_{-\infty}^0\df\lambda \int\limits_{-\infty}^0\df\lambda'\int\frac{\df^Dl}{(2\pi)^D}e^{-il\cdot(x_0+\lambda n_+-\lambda'n_-)}\delta(l^2)\theta(l^0)e^{-i\lambda \delta_+-i\lambda'\delta_-}\notag\,\\
		&=& \frac{\alpha_s C_F}{2\pi}\frac{e^{\eps\gamma_E}\mu^{2\eps}}{\Gamma(1-\eps)}\int\limits_0^{\infty}\df l_-\int\limits_0^{\infty}\df l_+\frac{(l_-l_+)^{-\eps}\,e^{-i(l_++l_-) x^0/2}}{\big(l_++\delta_+\big)\big(l_--\delta_-\big)}\,\nonumber\\
		&=& \frac{\alpha_s C_F}{2\pi}\frac{e^{\eps\gamma_E}}{\Gamma(1-\eps)}\left[\Gamma(\eps)\Gamma(1-\eps)\left(\frac{\mu}{\delta_+}\right)^{\eps}-\frac{\Gamma(1-\eps)}{\eps}(i\mu x^0/2)^{\eps}\right]\notag\\
		&&\times\,\left[\Gamma(\eps)\Gamma(1-\eps)\left(\frac{\mu}{-\delta_-}\right)^{\eps}-\frac{\Gamma(1-\eps)}{\eps}(i\mu x^0/2)^{\eps}\right],
  \label{eq:LPsoftunsb}
\end{eqnarray}
In expressions such as the above, we always expand in $\delta_\pm$ 
before expanding in $\eps$, and keep the leading non-analytic dependencies 
of the form $\delta_\pm^\eps$, while dropping $O(\delta_\pm)$ terms.
For the mirror diagram, the result reads 
\begin{equation}
	\label{eq:LPsoftunsbmirror}
	\begin{aligned}
		& \frac{\alpha_s C_F}{2\pi}\frac{e^{\eps\gamma_E}}{\Gamma(1-\eps)}\left[\Gamma(\eps)\Gamma(1-\eps)\left(\frac{\mu}{\delta_+^*}\right)^{\eps}-\frac{\Gamma(1-\eps)}{\eps}(i\mu x^0/2)^{\eps}\right]\\
		&\hspace{2.85cm}\times\left[\Gamma(\eps)\Gamma(1-\eps)\left(\frac{\mu}{-\delta_-^*}\right)^{\eps}-\frac{\Gamma(1-\eps)}{\eps}(i\mu x^0/2)^{\eps}\right].
	\end{aligned}
\end{equation}
Due to the $\delta_\pm$-regulators, both are finite as $\eps\to0$.  We reorganize the sum of \eqref{eq:LPsoftunsb} and \eqref{eq:LPsoftunsbmirror} as 
\begin{align}
	\label{eq:LPsoftunsr}
		S_{q\bar{q},{\rm uns}}^{{\rm LP}, r}(x_0) &= \frac{\alpha_s C_F}{\pi}\Bigg\{ \frac{e^{-\eps\gamma_E}\Gamma(1-\eps)}{\eps^2}\,e^{2\eps L_{x_0}}	+ e^{\eps\gamma_E}\Gamma(1-\eps)\Gamma^2(\eps)\,\mbox{Re}\left[ \left(\frac{\mu}{-\delta_-}\right)^{\!\eps}\left(\frac{\mu}{\delta_+}\right)^{\!\eps}\right]\notag\\
		& - \frac{\Gamma(1-\eps)\Gamma(\eps)}{2\eps}e^{\eps L_{x_0}}\bigg[\left(\frac{\mu}{\delta_+}\right)^{\!\eps}+\left(\frac{\mu}{\delta_+^*}\right)^{\!\eps}+\left(\frac{\mu}{-\delta_-}\right)^{\!\eps}+\left(\frac{\mu}{-\delta_-^*}\right)^{\!\eps}\bigg]\Bigg\}\\
        &= \mathcal{O}(\eps^0)\notag. 
\end{align}

While this expression and \eqref{eq:LPsoftunsv} are still valid for complex $\delta_\pm$ in the upper complex half-plane,\footnote{In writing \eqref{eq:LPsoftunsv}, we replaced $(-i\delta_{-})^{-\eps}(i\delta_+^*)^{-\eps}$ by $(\delta_{-})^{-\eps}(\delta_+^*)^{-\eps}$, implicitly defining the virtual correction with branch cuts along the negative real axis 
in the complex $\delta_\pm$-planes rather than in the imaginary direction. This guarantees that the virtual and real correction have the same analyticity domains and can be combined into an analytic function in the upper complex $\delta_\pm$-planes.} 
from now on we shall assume that $\delta_\pm$ are real with implicit $+i0^+$ as already indicated by the definitions \eqref{eq:delta}, \eqref{eq:deltaadj} of the regularized Wilson lines. 
After some algebra, the sum of the virtual and real corrections can be simplified to
\begin{align}
	\label{eq:LPsoftuns}
		S_{q\bar{q},{\rm uns}}^{{\rm LP}}(x_0) = 1 &+ \frac{\alpha_s C_F}{\pi} \Bigg\{ \frac{e^{-\eps\gamma_E}\Gamma(1-\eps)}{\eps^2}\,e^{2\eps L_{x_0}}	- \frac{\Gamma(1-\eps)\Gamma(\eps)}{2\eps}e^{\eps L_{x_0}}\left[\left(\frac{\mu}{\delta_+}\right)^\eps +\left(\frac{\mu}{\delta_+^*}\right)^\eps \right.\notag\\
        & \left.+\left(\frac{\mu}{-\delta_-}\right)^\eps+\left(\frac{\mu}{-\delta_-^*}\right)^\eps\,\right]- \frac{\pi^2}{2}\frac{e^{\eps\gamma_E}}{\Gamma(1-\eps)} \left(\frac{\mu^2}{|\delta_+\delta_-|}\right)^\eps \left[\cos\left(\frac{\pi\eps}{2}\right)\right]^{-2} \notag\\
       &\times{\rm sgn}\big({\rm Re}(\delta_+)\big){\rm sgn}\big({\rm Re}(\delta_-)\big) 
       \Bigg\}+\mathcal{O}(\alpha_s^2)\\
       =  1 &- \frac{\alpha_sC_F}{\pi}\left[\frac{1}{\eps^2}+\frac{1}{\eps}\left(\ln\frac{\mu}{|\delta_-|}+\ln\frac{\mu}{|\delta_+|}\right)+\mathcal{O}(\eps^0)\right] + \mathcal{O}(\alpha_s^2)\,.\notag
\end{align}
Clearly, the first term in the curly brackets of \eqref{eq:LPsoftuns} is nothing but \eqref{eq:LPsoft}, while the remaining terms depend on the $\delta$-regulators. The last term of the expression in curly brackets contributes a $\pi^2$ term, the sign of which depends on the relative sign of the $\delta$-regulators, and which arises  from an incomplete cancellation between \eqref{eq:LPsoftunsv} and the second term in \eqref{eq:LPsoftunsr}. This term is finite as $\eps\to 0$, and does not contribute to the anomalous dimension.  As an aside, we remark that the $\pi^2$ would be absent, if we had chosen $\delta_\pm$ to be purely imaginary rather than real. 

\subsection{Soft subtraction}
\label{subsect:subtraction}

The UV poles of \eqref{eq:LPsoftuns} depend on the $\delta$-regulators, which implies that the unsubtracted soft operator does not have a well-defined anomalous dimension. It must be divided by suitable subtraction factors that eliminate the $\delta$-dependence, which in turn must be multiplied back to collinear functions rendering both, soft and collinear functions, IR-finite and renormalizable.

%%%%%%%%%%%%%%%%%%%%%%%%%%%%%%%%%%%%%%%%%%%%%%%%%%%%%%%%%%%%%%%%%%%%%%%%%%
\begin{figure}[t]
\centering
\includegraphics[width=13cm]{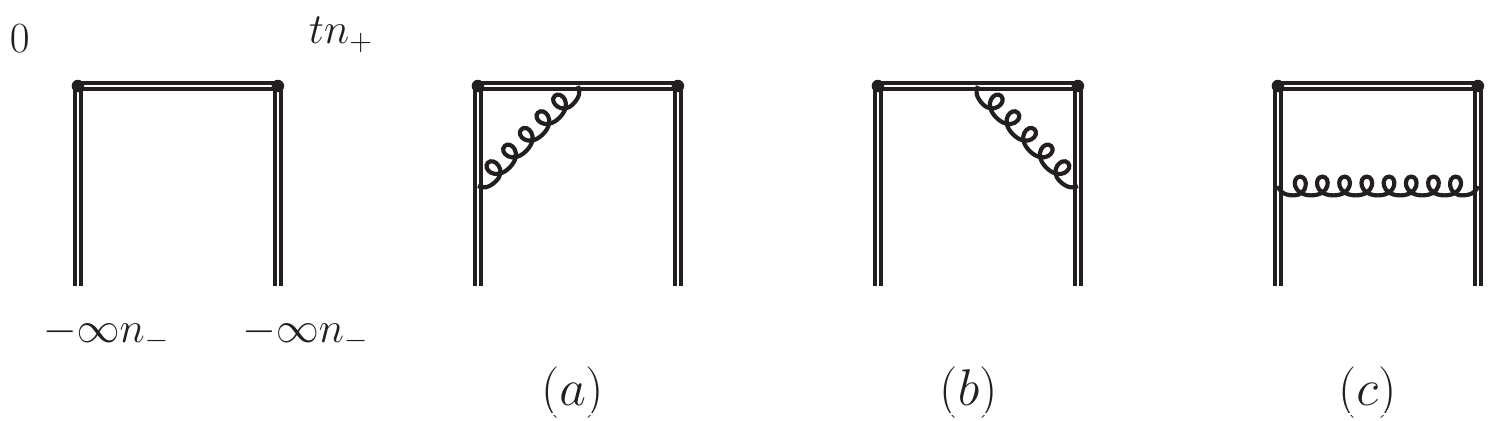}
\caption{\label{fig:LP-sub}The subtraction operator and its one-loop corrections. Diagram $(c)$ does not contribute in the Feynman gauge.
}
\end{figure}
%%%%%%%%%%%%%%%%%%%%%%%%%%%%%%%%%%%%%%%%%%%%%%%%%%%%%%%%%%%%%%%%%%%%%%%%%%
  
In the present case, we define the subtraction operators in the position space via 
\begin{equation}
	\label{eq:Wdef}
	\begin{aligned}
	        W_{\sqcap}^{-}(t) &\equiv \frac{1}{N_c}\,\mbox{Tr} \, \mathbf{T}\left[Y^\dagger_{n_-}(tn_+)\, [tn_+, 0]\, Y_{n_-}(0)\right],\\
        W_{\sqcap}^{+}(t) &\equiv \frac{1}{N_c}\,\mbox{Tr} \, \mathbf{T}\left[Y^\dagger_{n_+}(0)\, [0, tn_-]\, Y_{n_+}(tn_-)\right].
	\end{aligned}
\end{equation}
They are motivated by the factorization of quark and anti-quark parton distribution functions in the threshold limit $x\to 1$ \cite{Korchemsky:1992xv,Becher:2006mr,Falcioni:2019nxk}, where $x$ here denotes the momentum fraction, not the position. The subtractions are defined as the vacuum matrix element of these operators:
\begin{equation}
    \label{eq:Subdef}
    S_{\sqcap}^{\pm}(t) = \langle 0 |W_{\sqcap}^{\pm}(t) | 0 \rangle.
\end{equation}
We note that the subtraction defined here is almost identical to 
the one chosen in Appendix~B of~\cite{Chay:2013zya} for the off-shell regulated collinear functions 
in $x\to 1$ deep-inelastic scattering, with the difference that we apply time-ordering in \eqref{eq:Wdef}, which avoids the spurious IR divergence from the separation of the real and the virtual cut of diagrams $(a)$ 
and $(b)$. As discussed in App.~\ref{appendix:subopandPDF} here, it is not accidental that the same subtraction appears in the soft and collinear functions -- it is the very essence of the IR rearrangement that renders the renormalization of the soft and collinear functions separately well-defined. 

The one-loop corrections for $S_{\sqcap}^-(t)$ are shown in Fig.~\ref{fig:LP-sub}.
The semi-infinite Wilson lines are supplied with $\delta$-regulators as above. The straightforward calculation of the $\mathcal{O}(\alpha_s)$ correction leads to 
\begin{eqnarray}
        S_{\sqcap}^{-}(t) &=& 1-\frac{\alpha_s C_F}{2\pi}e^{\eps\gamma_E}\mu^{2\eps}\Gamma(1-\eps)\int\limits_0^t \df v \int\limits_{-\infty}^0\df u\bigg[\overbrace{e^{-iu \delta_-}(uv)^{-1+\eps}}^{(a)}+\overbrace{e^{iu \delta_-^*}(-uv)^{-1+\eps}}^{(b)}\bigg]\nonumber\\
        &=& 1-\frac{\alpha_s C_F}{2\pi} \frac{\Gamma(1-\eps)\Gamma(\eps)}{\eps}e^{\eps L_{ t}}\left[\left(\frac{\mu}{-\delta_-}\right)^{\!\eps}+\left(\frac{\mu}{-\delta_-^*}\right)^{\!\eps}\,\right]\nonumber\\
        &=& 1 - \frac{\alpha_sC_F}{\pi}\left[\frac{1}{\eps^2}+\frac{1}{\eps}\left(L_t+\ln\frac{\mu}{|\delta_-|}\right)+\mathcal{O}(\eps^0)\right]\,,
     \label{eq:LP-subm}
\end{eqnarray}
where $L_t$ is defined as in \eqref{eq:Lx0} with replacement $x^0/2\to t$. Similarly
\begin{equation}
    \label{eq:LP-subp}
    \begin{aligned}
        S_{\sqcap}^{+}(t) &= 1-\frac{\alpha_s C_F}{2\pi} \frac{\Gamma(1-\eps)\Gamma(\eps)}{\eps}e^{\eps L_{t}}\left[\left(\frac{\mu}{\delta_+}\right)^{\!\eps}+\left(\frac{\mu}{\delta_+^*}\right)^{\!\eps}\,\right]\\
        &= 1 - \frac{\alpha_sC_F}{\pi}\left[\frac{1}{\eps^2}+\frac{1}{\eps}\left(L_t+\ln\frac{\mu}{|\delta_+|}\right)+\mathcal{O}(\eps^0)\right].
    \end{aligned}
\end{equation}

Combining \eqref{eq:LPsoftuns}, \eqref{eq:LP-subm} and \eqref{eq:LP-subp}, 
one finds that the $\delta_\pm$-regulators can be removed in the 
subtracted soft function defined as 
\begin{align}
    \label{eq:subdef}
    S_{q\bar{q}, {\rm sub}}^{\rm LP}(x_0) \equiv  \frac{S_{q\bar{q},{\rm uns}}^{\rm LP}(x_0)}{S_{\sqcap}^{+}(x^0/2)S_{\sqcap}^{-}(x^0/2)} \stackrel{\text{pole}}{=} S_{q\bar{q}}^{\rm LP}(x_0). 
\end{align}
Note that the second equality holds only for the pole part in $\epsilon$, which is enough to extract the anomalous dimension. 

We observe that the above subtractions cancel exactly the second term in curly brackets in \eqref{eq:LPsoftuns} with full $\eps$ dependence, leaving
\begin{eqnarray}
	S_{q\bar{q},{\rm sub}}^{{\rm LP}}(x_0) &=& S_{q\bar{q}}^{\rm LP}(x_0) - \frac{\alpha_s C_F}{\pi} \frac{\pi^2}{2}\frac{e^{\eps\gamma_E}}{\Gamma(1-\eps)} \left(\frac{\mu^2}{|\delta_+\delta_-|}\right)^\eps \left[\cos\left(\frac{\pi\eps}{2}\right)\right]^{-2} \notag\\[0.2cm]
&& \times\,{\rm sgn}\big({\rm Re}(\delta_+)\big){\rm sgn}\big({\rm Re}(\delta_-)\big) +\mathcal{O}(\alpha_s^2).
\end{eqnarray}
The subtracted soft function is not identical to the soft function computed in dimensional regularization for the UV {\em and} IR divergences.  
The mismatch is a finite term proportional to $\pi^2$. Such a mismatch was also noted in \cite{Chay:2013zya}, where a similar subtraction in the momentum space was used. 
We do not pursue the origin of this finite term here (although it might be interesting), since we employ the $\delta$-regulators only as a means to extract the divergent parts / anomalous dimension, but note again that the mismatch would be absent for purely imaginary $\delta$-regulators.

The subtractions $S_{\sqcap}^{\pm}(t)$ must be {\em multiplied} to the $\delta$-regulated light-cone operators that define the parton distributions, which are convoluted with the soft function in the factorization theorem for the DY process near threshold. We demonstrate the consistency of this procedure in App.~\ref{appendix:subopandPDF}, 
that is, that the UV divergence of the $\delta$-regulated parton distributions are given by the DGLAP kernel in the $x\to 1$ limit. 

\subsection{Anomalous dimension}
\label{sect:RGELP}
In position space the LP soft function is renormalized multiplicatively to all orders \cite{Korchemskaya:1992je}. Hence, we can determine the renormalization factor in the $\overline{\text{MS}}$ scheme from \eqref{eq:LPsoft} by 
requiring 
\begin{equation}
	\label{eq:LPsoftrenor}
	S_{q\bar{q}}^{\rm LP}(x_0;\mu) = Z_S^{\rm LP} S_{q\bar{q}}^{\rm LP}(x_0)\stackrel{!}{=} \mbox{finite} =1 + \frac{\alpha_s C_F}{\pi } \left[ 2L_{x_0}^2+\frac{\pi^2}{12}\right]+\mathcal{O}(\alpha_s^2),
\end{equation}
for the renormalized soft function $S_{q\bar{q}}^{\rm LP}(x_0;\mu)$, which yields
\begin{equation}
	\label{eq:LPsoftrenorZS}
	 Z_S^{\rm LP}(x_0) = 1 + \frac{\alpha_s C_F}{\pi} \left[-\frac{1}{\eps^2}-\frac{2L_{x_0}}{\eps}\right]+\mathcal{O}(\alpha_s^2).
\end{equation}
The position-space anomalous dimension in the $\overline{\text{MS}}$ scheme reads
\begin{equation}
	\label{eq:LPsoftAD}
	\gamma_S^{\rm LP}(x_0) \equiv - \frac{\df\,Z_S^{\rm LP}(x_0)}{\df \ln\mu}\frac{1}{Z_S^{\rm LP}(x_0)}= -4\frac{\alpha_s C_F}{\pi} L_{x_0}+\mathcal{O}(\alpha_s^2)\,,
\end{equation}
consistent with the all-order form
\begin{equation}
	\label{eq:LPsoftADall}
	\gamma_S^{\rm LP}(x_0) = -4\Gamma_{\rm cusp}^{\rm F}(\alpha_s)L_{x_0}+2 \gamma_W(\alpha_s),
\end{equation}
in terms of the cusp anomalous dimension in the fundamental representation (see App.~\ref{appendix:RG}) and $\gamma_W(\alpha_s)=\mathcal{O}(\alpha_s^2)$. The renormalization-group equation (RGE) 
\begin{equation}
	\label{eq:LPsoftRGE}
	\frac{\rm d}{{\rm d}\ln\mu} S_{q\bar{q}}^{\rm LP}(x_0;\mu) = - \gamma_S^{\rm LP}(x_0) \,S_{q\bar{q}}^{\rm LP}(x_0;\mu)
\end{equation}
and its solution are also multiplicative.

Employing \eqref{eq:FourierOmega} and writing $L_{x_0}$ in terms of $\df/\df\eta$, the anomalous dimension in momentum-space is found to be
\begin{equation}
	\label{eq:LPsoftADmom}
	\gamma_S^{\rm LP}(\Omega,\Omega') = \left[4\Gamma_{\rm cusp}^{\rm F}(\alpha_s)\ln\frac{\Omega}{\mu}+2\gamma_W(\alpha_s)\right]\delta(\Omega-\Omega')+4\Gamma_{\rm cusp}^{\rm F}(\alpha_s)\,\Omega\,\left[\frac{\theta(\Omega-\Omega')}{\Omega\left(\Omega-\Omega'\right)}\right]_+\!,
\end{equation}
where the plus distribution acting on a test function $g(\Omega')$ is defined as
\begin{equation}
	\label{eq:LNdef}
	\int\limits_0^\infty\df\Omega'\left[f(\Omega,\Omega')\right]_+g(\Omega') = \int\limits_0^\infty\df\Omega'f(\Omega,\Omega')\left( g(\Omega') -g(\Omega) \right). 
\end{equation}

% -----------------------------------------------------------------------------

\section{Off-diagonal DY NLP soft-quark function}
\label{sect:softNLP}

The  soft operator for the power-suppressed $g\bar{q}$ DY production channel is defined as\footnote{The definition differs from the one in \cite{Broggio:2023pbu} by a colour factor 2 and by removing an 
inverse derivative $1/(i\nm\partial)$ from the $\mathbf{T}$ and $\mathbf{\bar{T}}$ product, which is more conveniently 
assigned to the collinear function as done for the corresponding soft function for ``gluon thrust''~\cite{Beneke:2022obx}. }
\begin{eqnarray}
    \label{eq:NLPsoftoperatordef1}
	O_{g\bar{q},{\rm uns}}^{{\rm NLP}}(x_0, \{s\}) &\equiv& \frac{g_s^2}{N_c C_F}\,\mbox{Tr} \, \mathbf{\bar{T}}\left[ \big[\bar{q}_sY_{n_-}\big](x_0+s_2 n_-) T^a Y^\dagger_{n_-}(x_0) \,Y_{n_+}(x_0)\right] \,\frac{\slashed{n}_-}{4}\notag \\
	&& \hspace{0.0cm} \times\mathbf{T} \left[ Y^\dagger_{n_+}(0)Y_{n_-}(0) T^a\big[Y_{n_-}^\dagger q_s\big](s_1n_-) \right].
\end{eqnarray}
where $\{s\}$ collectively denotes $(s_1, s_2)$. Likewise the corresponding momentum-space variables will be denoted by $\{\omega\}$ later on. The trace here is with respect to the colour indices and will take the Dirac indices into account as well when taking the vacuum matrix element of the operator to arrive at the soft function.
We combine the two $\nm$-directed Wilson lines in the fundamental representation using the identity 
\begin{equation}
   \begin{aligned}
       Y_{\nm}(x+t_1 \nm) T^a Y_{\nm}^\dagger(x+t_2 \nm) &= \mathcal{Y}^{ca}_{n_-} (x+t_1\nm)\,T^c\, [x+t_1\nm, x+t_2 \nm]  \\
       &= [x+t_1\nm, x+t_2 \nm]\,T^c\,\mathcal{Y}^{ca}_{n_-} (x+t_2\nm),
   \end{aligned}  
   \label{eq:wrel}
\end{equation}
where the finite-distance Wilson line is defined as in \eqref{eq:Wilsonlength}, 
to rewrite the soft operator as
\begin{eqnarray}
    \label{eq:NLPsoftoperatordefnew}
    O_{g\bar{q},{\rm uns}}^{{\rm NLP}}(x_0, \{s\})&=& 
    \frac{g_s^2}{N_c C_F}\,\mbox{Tr} \, \mathbf{\bar{T}}\left[ \bar{q}_s(x_0+s_2n_-) (\mathcal{Y}^\dagger)^{ac}_{n_-}(x_0+s_2n_-)T^c\left[x_0+s_2 n_-, x_0\right] Y_{n_+}(x_0)\right]  \notag\\
	&& \hspace{0.0cm}\times\frac{\slashed{n}_-}{4}\mathbf{T} \left[ Y^\dagger_{n_+}(0)[0,s_1n_{-}]T^d\,\mathcal{Y}^{da}_{n_-}(s_1n_-)\,q_s(s_1n_-) \right].
\end{eqnarray}
The vacuum matrix element of the operator $O_{g\bar{q},{\rm uns}}^{{\rm NLP}}(x_0, \{s\})$ gives the position-space soft function, $S_{g\bar{q}, \rm uns}^{{\rm NLP}}(x_0, \{s\})$. 

The form of \eqref{eq:NLPsoftoperatordefnew} is best adapted to the physical process $g+\bar{q}\to \gamma^*+X$, 
and reflects the classical space-time picture of soft radiation from the moving colour charges. Focusing on the $\mathbf{T}$-product, the adjoint Wilson line $\mathcal{Y}^{da}_{n_-}(s_1n_-)$ describes the soft radiation from the energetic gluon incoming from $-\infty$ to $s_1 n_-$, at which point it emits a soft anti-quark $q_s(s_1n_-)$ and turns into an energetic quark that travels a finite light-like distance from $s_1 n_-$ to the hard interaction point 0 
(Wilson line $\left[0, s_1 n_-\right]$), where it annihilates with the incoming energetic anti-quark. The semi-infinite Wilson line $Y^\dagger_{n_+}(0)$ 
accounts for the radiation from this anti-quark. 
(In the complex-conjugated amplitude, all points are shifted by the time-like vector $x_0$.).

We remark that unless the $\delta$-regulators are purely imaginary, the regulated Wilson lines are not unitary and the adjoint Wilson line \eqref{eq:deltaadj} is not real. As a consequence, \eqref{eq:wrel} and $\mathcal{Y}^\dagger=\mathcal{Y}^T$ do not hold, and \eqref{eq:NLPsoftoperatordef1}, \eqref{eq:NLPsoftoperatordefnew}, as well as \eqref{eq:NLPsoftoperatordefnew} with $\mathcal{Y}^T$ in the anti-time-ordered part all define different regulated soft functions. We shall comment below on the differences this entails for the result of the one-loop computations. 

\subsection{Leading order}
The leading-order soft-function is obtained by setting all Wilson lines to 1, and evaluating the cut quark propagator, resulting in
\begin{align}
   \label{eq:OgqLO}
   S_{g\bar{q}}^{{\rm NLP}}(x_0, \{s\}) &= g_s^2\int\frac{\df^D p}{(2\pi)^{D-1}}e^{-ix_0\cdot p +i(s_1-s_2)n_-\cdot p}\,\mbox{Tr} \left[\slashed{p}\frac{\slashed{n}_-}{4}\right]\delta(p^2)\theta(p^0)\notag \\
   &= \frac{\alpha_s}{4\pi}e^{\eps\gamma_E}\Gamma(2-\eps)\mu^{2\eps}\left[i\left(\frac{x^0}{2}+s_2-s_1-i0^+\right)\right]^{-2+\eps}\left[i\frac{x^0}{2}\right]^{-1+\eps}\,.
\end{align} 

In the following, we take Fourier transformations to obtain the momentum-space soft function. 
Since later we will show that the $x^0$ dependence is multiplicative as for the LP soft function, 
we first perform the Fourier transforms only with respect to $s_1$, $s_2$ to obtain the ``mixed-space'' soft function. 
Changing variable from $s_2$ to $\delta s = s_2-s_1$ in \eqref{eq:OgqLO}, and integrating over $s_1$ results in  $\delta(\omega_1-\omega_2)$ and the expression 
\begin{align}
   \label{eq:SgqLOpartialF}
   S^{\rm NLP}_{g\bar{q}}(x_0,\{\omega\}) &= \int\limits^{+\infty}_{-\infty} \df s_1 \int\limits^{+\infty}_{-\infty} \df s_2\, e^{-i\omega_1s_1+i\omega_2s_2} \,S_{g\bar{q}}^{{\rm NLP}}(x_0, \{s\})
   \notag\\ 
   &=\frac{\alpha_s}{4\pi} e^{\eps\gamma_E}\mu^{2\eps}\delta(\omega_1-\omega_2)\theta(\omega_1)\omega_1^{1-\eps} e^{-i \omega_1 x^0/2}\left[i\frac{x^0}{2}\right]^{\eps-1}
\end{align}
for the lowest-order mixed-space soft function. The momentum-space soft function is then obtained  by the further Fourier transform with respect to $x^0$, resulting in 
\begin{eqnarray}
    \label{eq:SgqLO}
	S^{\rm NLP}_{g\bar{q}}(\Omega, \{\omega\}) &=& \int\frac{\df x^0}{4\pi}\,e^{i\Omega x^0/2}\,S_{g\bar{q}}^{{\rm NLP}}(x_0, \{\omega\})\\
    &=& \frac{\alpha_s}{4\pi} e^{\eps\gamma_E}\mu^{2\eps}\delta(\omega_1-\omega_2)\theta(\omega_1)\omega_1^{1-\eps}\int\frac{\df x^0}{4\pi} e^{i(\Omega-\omega_1) x^0/2}\left[i\frac{x^0}{2}\right]^{\eps-1} \notag\\
    &=& \frac{\alpha_s}{4\pi}\frac{e^{\eps\gamma_E}}{\Gamma(1-\eps)}\omega_1\left(\frac{\mu^2}{\omega_1(\Omega-\omega_1)}\right)^{\!\eps}\theta(\omega_1)\theta(\Omega-\omega_1)\delta(\omega_1-\omega_2)\,.   
    \label{eq:SgqLOres}
 \end{eqnarray}
Due to the $i0^+$ prescriptions of the position coordinates, the momentum variables have positive support. As a result, the Fourier inversion reads
\begin{align}
\label{eq:renormpm}
   S_{g\bar{q}}^{{\rm NLP}}(x_0, \{s\}) = \int\limits_0^\infty\df \Omega\int\limits_0^\infty\{\df\omega\}\,e^{-i\Omega (x^0-i0^+)/2+i\omega_1 (s_1+i0^+)-i\omega_2 (s_2-i0^+)} S_{g\bar{q}}^{{\rm NLP}}(\Omega, \{\omega\}),
\end{align}
for {\it both} the bare and renormalized functions, and $\{\df\omega\}$ denotes two-folds integrations over $\omega_1$ and $\omega_2$. In the above, we made the $i0^+$ prescription explicit.

\subsection{Next-to-leading order, unsubtracted}
\label{sect:softNLPuns}

We next turn to the soft operator at next-to-leading order (NLO). We begin with extracting UV poles in the dimensional regulator $\eps$ by introducing $\delta_\pm$  to regulate the IR singularities. Then, we follow the same logic as in the LP case and introduce subtracted 
operators.  Unlike the LP case, we only extract the UV poles in $\eps$, since we are interested only in the renormalization factor. The soft function at NLO with 
dimensional regularization for UV {\em and} IR singularities has already been computed in 
\cite{Broggio:2023pbu}.

%%%%%%%%%%%%%%%%%%%%%%%%%%%%%%%%%%%%%%%%%%%%%%%%%%%%%%%%%%%%
\begin{figure}[t]
\centering
\includegraphics[width=14cm]{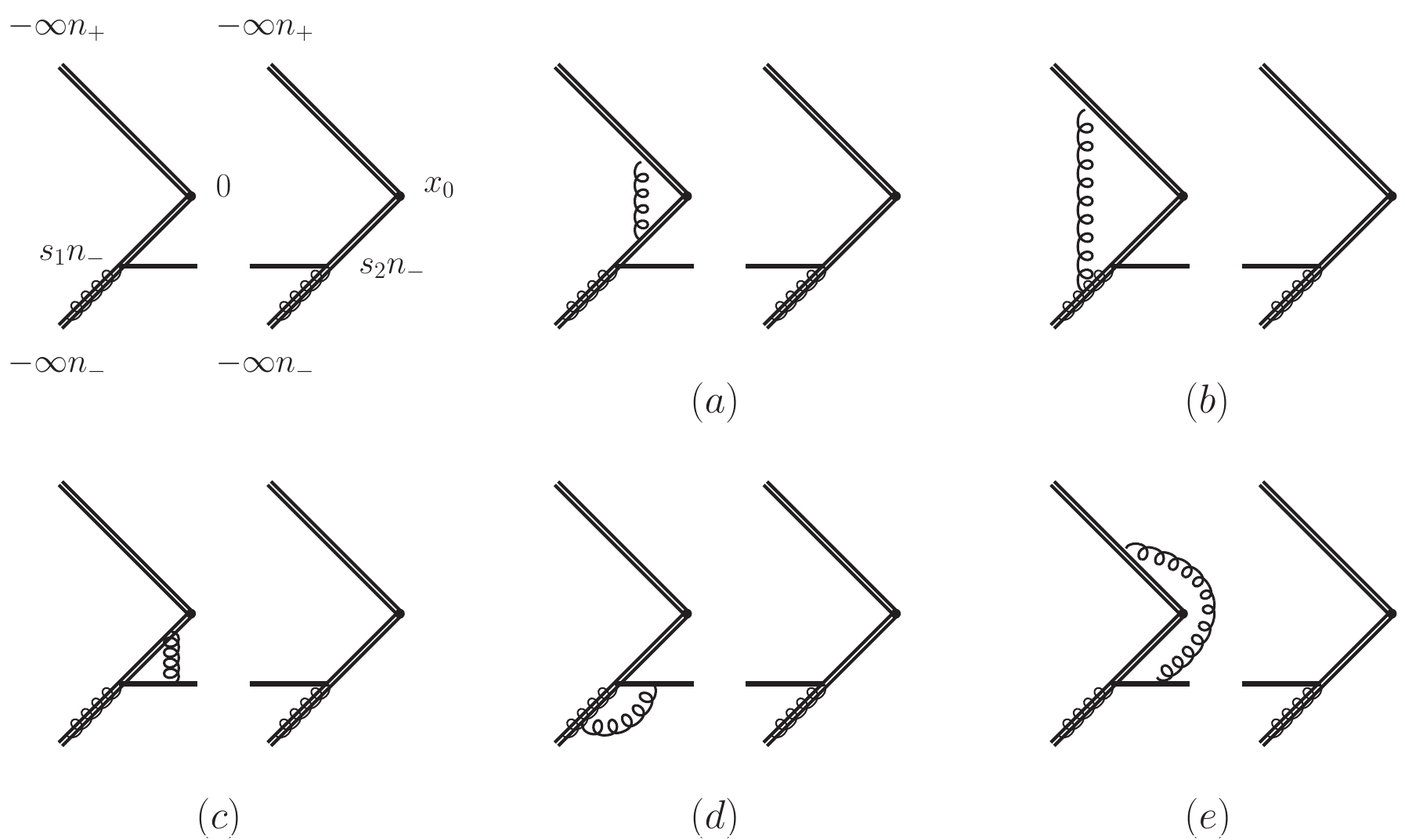}
\caption{\label{fig:NLP-virtual}The soft function entering the off-diagonal Drell-Yan process and its one-loop virtual corrections. The mirror ones are not shown. Diagram $(e)$ does not have UV poles in $\eps$, but we put it here for completeness. Solid lines are quark and anti-quark fields; double lines are Wilson lines in the fundamental representation, while double lines with overlapped gluon lines are Wilson lines in the adjoint representation.}
\end{figure}
%%%%%%%%%%%%%%%%%%%%%%%%%%%%%%%%%%%%%%%%%%%%%%%%%%%%%%%%%%%%%

\begin{figure}[t]
\centering
\includegraphics[width=11cm]{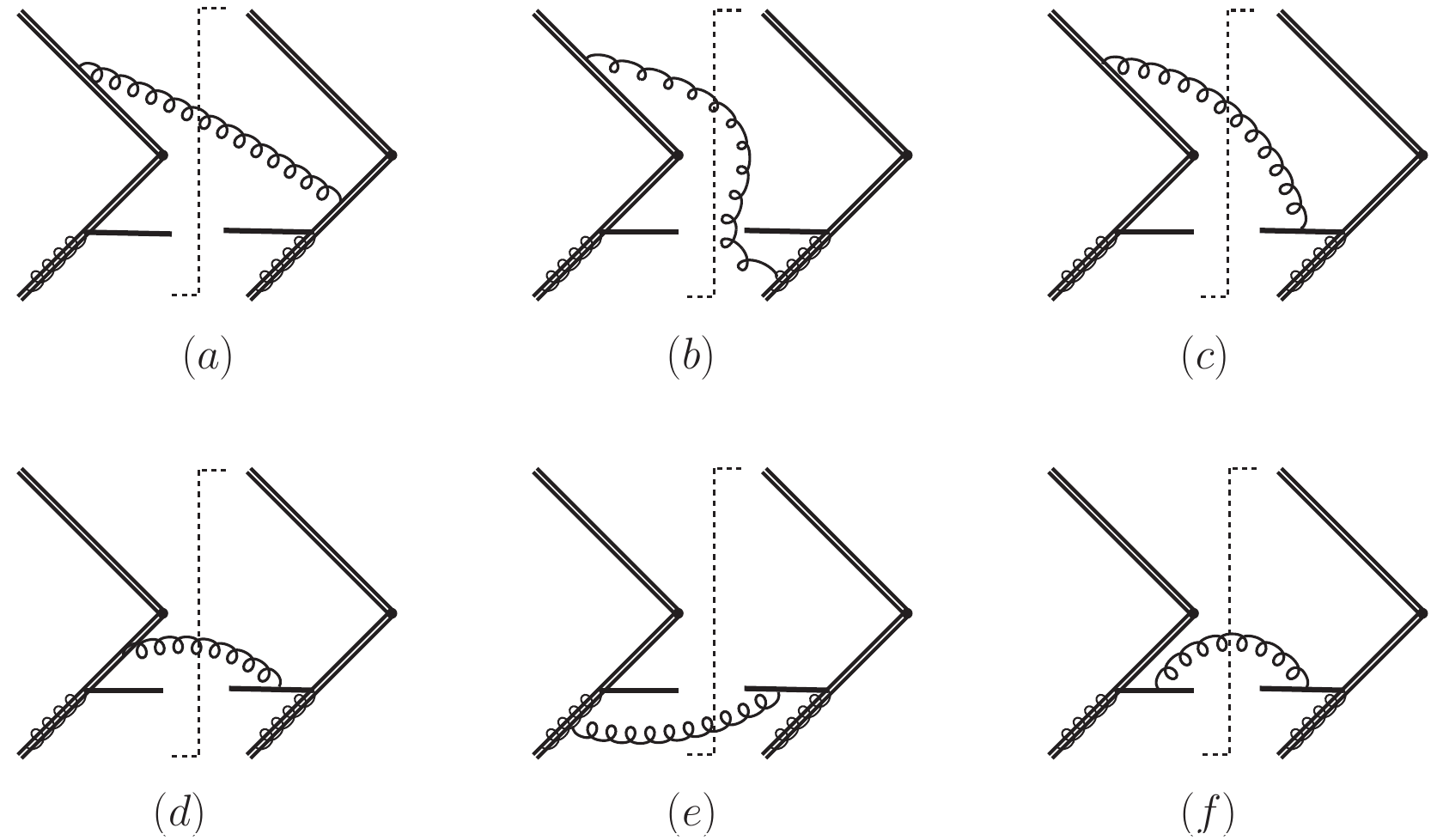}
\caption{\label{fig:NLP-real}One-loop real-emission corrections. The dashed line denotes the cut. The mirror diagrams are not shown. Neither are correlations between Wilson lines in the same direction, which are trivially zero in Feynman gauge. }
\end{figure}

We use the background-field technique~\cite{Balitsky:1987bk} as in \cite{Beneke:2024cpq} and present results in  Feynman gauge. Calculations in general $R_\xi$ gauge are discussed in App.~\ref{appendix:gauge}.  The one-loop virtual and real corrections are shown in Figs.~\ref{fig:NLP-virtual} and~\ref{fig:NLP-real}, where mirror diagrams are not shown explicitly.

Turning to the virtual corrections first, we start with the cusp diagram $(a)$.
We find 
\begin{align}
	\label{eq:NLPsoftva}
	I_a^{\left(\delta_+^*\right)} &=  -ig_s^2 2C_F\int\limits_0^{s_1}\df\lambda\int\limits_{-\infty}^0\df\lambda' e^{i\lambda'\delta_+^*}\int\frac{\df^D l}{(2\pi)^D}\frac{e^{-il\cdot(\lambda' n_+-\lambda n_-)}}{-l^2-i0^+}O_{g\bar{q},{\rm uns}}^{{\rm NLP}}(x_0,\{s\})\notag\\
	&= \frac{\alpha_s}{4\pi}2C_Fe^{\eps\gamma_E}\Gamma(1-\eps)\mu^{2\eps}\int\limits_0^{s_1}\df\lambda\int\limits_{-\infty}^0\df\lambda' e^{i\lambda'\delta_+^*}\left(\lambda\lambda'\right)^{-1+\eps}O_{g\bar{q},{\rm uns}}^{{\rm NLP}}(x_0,\{s\})\notag\\
	&= \frac{\alpha_s}{4\pi}2C_F\left[-\frac{1}{\eps^2}-\frac{1}{\eps}\ln\frac{\mu}{\delta_+^*}-\frac{1}{\eps}\ln\left(i\mu e^{\gamma_E}s_1\right)\right]O_{g\bar{q},{\rm uns}}^{{\rm NLP}}(x_0,\{s\})+\mathcal{O}(\eps^0)\,.
\end{align}
The superscript indicates the dependence on $\delta_+^*$.
Diagram $(b)$ has only a simple $1/\eps$ pole and reads
\begin{align}
	\label{eq:NLPsoftvb}
	I_b^{\left(\delta_-\right)} &=  ig_s^2 C_A\int\limits_{-\infty}^0 \df\lambda\int\limits_{-\infty}^0\df\lambda' e^{-i\lambda\delta_-+i\lambda'\delta_+^*}\int\frac{\df^D l}{(2\pi)^D}\frac{e^{-il\cdot(\lambda' n_+-(s_1+\lambda) n_-)}}{-l^2-i0^+}O_{g\bar{q},{\rm uns}}^{{\rm NLP}}(x_0,\{s\})\notag\\
	&= \frac{\alpha_s}{4\pi}\frac{C_A}{\eps}\left[\ln\left(i\mu e^{\gamma_E}s_1\right)-\ln\frac{\mu}{\delta_-}\right]O_{g\bar{q},{\rm uns}}^{{\rm NLP}}(x_0,\{s\})+\mathcal{O}(\eps^0).
\end{align}
The superscript indicates the dependence on $\delta_-$. The expressions for the corresponding two mirror diagrams are obtained by replacing $\delta_\pm \to \delta^*_\pm$ and $i s_1\to -i s_2$.

Diagrams $(c)$ and $(d)$ involve correlations between the quark field and the Wilson lines. They are the same as in \cite{Beneke:2024cpq} and given by
\begin{align}
	\label{eq:NLPsoftvc}
	I_c = \frac{\alpha_s}{4\pi}\left(C_F-\frac{C_A}{2}\right)\frac{2}{\eps}\int\limits_0^1\df u\left[\frac{u}{\bar{u}}\right]_+O_{g\bar{q},{\rm uns}}^{{\rm NLP}}(x_0,u\,s_1, s_2)+\mathcal{O}(\eps^0),
\end{align}
where $\bar{u}=1-u$ and the plus distribution in this context is defined as
\begin{equation}
	\label{eq:plusdist}
	\int\limits_0^1\df u\left[\frac{u}{\bar{u}}\right]_+f(u) = \int\limits_0^1\df u \frac{u}{\bar{u}}\left[ f(u)-f(1) \right].
\end{equation}
The mirror diagram of $(c)$ reads
\begin{align}
	\label{eq:NLPsoftvcmirror}
	\frac{\alpha_s}{4\pi}\left(C_F-\frac{C_A}{2}\right)\frac{2}{\eps}\int\limits_0^1\df u\left[\frac{u}{\bar{u}}\right]_+O_{g\bar{q},{\rm uns}}^{{\rm NLP}}(x_0,s_1, u\,s_2)+\mathcal{O}(\eps^0)\,.
\end{align}
Diagram $(d)$ requires $\delta_-$ again. Its result reads
\begin{align}
	\label{eq:NLPsoftvd}
	I_d^{\left(\delta_-\right)} = \frac{\alpha_s}{4\pi}\frac{C_A}{\eps}\left[1-\ln\frac{i\partial_{s_1}}{\mu}-\ln\frac{\mu}{\delta_-}\right]O_{g\bar{q},{\rm uns}}^{{\rm NLP}}(x_0, \{s\})+\mathcal{O}(\eps^0),
\end{align}
and its mirror is obtained by substituting  $\delta_-\to\delta_-^*$, 
$i\partial_{s_1}\to -i\partial_{s_2}$.

Diagram $(e)$ does not contain pole terms in $\eps$, since it correlates the quark field at $s_1 n_-$ with a Wilson line in a different light-like direction $n_+$, which tames its UV behaviour. Details of the argument can be found in \cite{Beneke:2024cpq}.  
In total, the one-loop virtual corrections read
\begin{align}
	\label{eq:NLPsoftv}
	I_{{\rm uns}, \text{virt}}^{\left(\delta_\pm\right)} =\,& \frac{\alpha_s}{2\pi\eps}\left(C_F-\frac{C_A}{2}\right)\int\limits_0^1\df u\left[\frac{u}{\bar{u}}\right]_+\Big[O_{g\bar{q},{\rm uns}}^{{\rm NLP}}(x_0,u\,s_1, s_2)+O_{g\bar{q},{\rm uns}}^{{\rm NLP}}(x_0,s_1, u\,s_2)\Big]\\
	 &+\frac{\alpha_s}{4\pi\eps} \Bigg\{ C_A\left[\ln\left(\mu^2 e^{2\gamma_E}s_1s_2\right)-2\ln\frac{\mu^2}{\left|\delta_-\right|^2} -\ln\frac{\partial_{s_1}\partial_{s_2}}{\mu^2}+2\right]\notag\\
	&-C_F\left[\frac{4}{\eps}+2\ln\frac{\mu^2}{\left|\delta_+\right|^2}+2\ln\left(\mu^2 e^{2\gamma_E}s_1s_2\right)\right] \Bigg\}\,O_{g\bar{q},{\rm uns}}^{{\rm NLP}}(x_0, \{s\})+\mathcal{O}(\eps^0),\notag
\end{align}
where $\ln(\mu/\delta_\pm)+\ln(\mu/\delta_\pm^*)=\ln(\mu^2/\left|\delta_\pm\right|^2)$ has been used. The virtual correction does not involve non-trivial dependence on the time-like displacement vector $x_0$, just like the leading-power function, but does depend on the $\delta_\pm$ regulators and is real-valued as expected. To obtain the $x_0$ dependence, we need to consider the real emission and corresponding subtractions, which we turn to in the following.

One-loop corrections for real emission are shown in Fig.~\ref{fig:NLP-real}. Mirror diagrams and diagrams where correlations are between Wilson lines in the same direction, which are trivially zero in Feynman gauge, are not shown. Like in the leading-power counterpart, diagrams $(a)$ and $(b)$ with $\delta_\pm$ regulators contribute only at $\mathcal{O}(\eps^0)$. Diagrams $(c)-(f)$ and their mirrors do not contribute either. The logic follows~\cite{Beneke:2024cpq}, so we do not give details here for brevity. As a result, we find
\begin{align}
	\label{eq:NLPsoftr}
	I_{{\rm uns}, \text{real}}^{\left(\delta_\pm\right)} = 0+\mathcal{O}(\eps^0).
\end{align}

A few remarks regarding the absence of a UV pole in the one-loop correction to the real emission are in order. First of all, we note that diagram $(a)$ in Fig.~\ref{fig:NLP-real} here is similar to diagram~$(b)$ in Figure 2 of~\cite{Beneke:2024cpq}, which does contribute a simple UV pole. 
The geometric reason is that the two light-like Wilson lines therein \textit{can} intersect at $x=0$, since $s$ and $t$ can take values in $]-\infty,\infty[$.  This is in contrast to the current case, where the two correlated Wilson lines in diagram $(a)$ never meet, regardless of the values of $s_1$ and $s_2$ due to the time-like displacement vector $x_0$. The same analysis also applies to diagram $(b)$ here, leading to the conclusion that the Wilson-Wilson line correlations in the real-emission diagrams are UV finite at the one-loop order.

%%%%%%%%%%%%%%%%%%%%%%%%%%%%%%%%%%%%%%%%%%%%%%%%%%%%%%%%%%
%%%%%%%%%%%%%%%%%%%%%%%%%%%%%%%%%%%%%%%%%%%%%%%%%%%%%%%%%%

In addition to the above one-loop corrections, the $\overline{\rm MS}$ quark field renormalization $q^{\mathrm{bare}} = Z_q^{1 / 2} q^{\mathrm{ren}}$ is required with 
\begin{equation}
	\label{eq:ZqZa}
		Z_q = 1-\frac{\alpha_s}{4 \pi} \frac{C_F}{\varepsilon}+\mathcal{O}\left(\alpha_s^2\right)
\end{equation} 
in Feynman gauge. Since the definition of the soft operator includes an overall $g_s^2$, the strong coupling renormalization is also necessary: 
\begin{equation}
 	\label{eq:Zas}
 	Z_{\alpha_s} = 1-\frac{\alpha_s}{4\pi}\frac{\beta_0}{\eps}+\mathcal{O}(\alpha_s^2),
 \end{equation}
with $\beta_0=11C_A/3-4T_Fn_f/3$ being the first coefficient of the QCD $\beta$-function, and $n_f$ the number of active quark flavours. Summing up all contributions, we obtain the one-loop UV poles of the unsubtracted operator~$O_{g\bar{q},{\rm uns}}^{\rm NLP}$,
\begin{eqnarray}
   \label{eq:NLPsoftuns}	
  O_{g\bar{q},{\rm uns}}^{\rm NLP,(1)}(x_0,\{s\}) &=& 
  O_{g\bar{q},{\rm uns}}^{\rm NLP,(0)}(x_0,\{s\}) 
	 +\frac{\alpha_s}{4\pi\eps} \Bigg\{ C_A\left[\ln\left(\mu^2 e^{2\gamma_E}s_1s_2\right)-2\ln\frac{\mu^2}{\left|\delta_-\right|^2} -\ln\frac{\partial_{s_1}\partial_{s_2}}{\mu^2}+2\right]\notag\\
	&&\hspace*{-2cm}-\,C_F\left[\frac{4}{\eps}+2\ln\frac{\mu^2}{\left|\delta_+\right|^2}+2\ln\left(\mu^2 e^{2\gamma_E}s_1s_2\right)+1\right]-\beta_0\Bigg\}\,
  O_{g\bar{q},{\rm uns}}^{{\rm NLP},(0)}(x_0, \{s\})\notag\\
	&&\hspace*{-2cm} +\, \frac{\alpha_s}{2\pi\eps}\left(C_F-\frac{C_A}{2}\right)\int\limits_0^1\df u\left[\frac{u}{\bar{u}}\right]_+ \Big[O_{g\bar{q},{\rm uns}}^{{\rm NLP},(0)}(x_0,u\,s_1, s_2)+O_{g\bar{q},{\rm uns}}^{{\rm NLP},(0)}(x_0,s_1, u\,s_2)\Big]\, ,
\end{eqnarray}
where the superscripts $(1)$ and $(0)$ denote the one-loop and tree-level bare NLP soft operator, respectively. Note that the $\delta_-$-dependent term is only relevant to the colour factor $C_A$, while the $\delta_+$-dependent term appears only with the colour factor $C_F$ as a consequence of the colour charges of the external partons. 

As mentioned above, in the presence of the $\delta$ regulators the Wilson lines are not unitary and the adjoint one is not real. We find that when the computation is done starting from \eqref{eq:NLPsoftoperatordef1}, the result coincides with \eqref{eq:NLPsoftuns}. On the other hand, employing  \eqref{eq:NLPsoftoperatordefnew} with $\mathcal{Y}^\dagger \to \mathcal{Y}^T$ 
in the anti-time ordered part implies that the mirror diagrams follow by 
substituting $\delta_-\to -\delta_-$ in \eqref{eq:NLPsoftvb} and 
\eqref{eq:NLPsoftvd}, which leads to $\ln\,(\mu^2/|\delta_-|^2) \to 
\ln\,(\mu^2/(-\delta_-^2))$ in \eqref{eq:NLPsoftv} and \eqref{eq:NLPsoftuns}. 
Hence, in the second case the corresponding {\em unsubtracted} soft function differs by an imaginary term.

% -----------------------------------------------------------------------------

\subsection{Subtraction}
\label{sect:softNLPsub}

The presence of the $\delta_\pm$ regulators in~\eqref{eq:NLPsoftuns} calls for the introduction of an IR subtraction. To understand the form of the subtraction, 
we recall that the factorization theorem for the hadronic DY process at NLP involves amplitude-level threshold- (or hard-) collinear functions and cross-section-level collinear functions, the parton distributions. The IR singularities of the soft functions must be connected to the latter, since the virtuality 
of the hard-collinear functions is large compared to the one of the soft 
function. This suggests that the subtraction is again related to the 
$x\to 1$ limit of parton distributions, analogous to the subtraction~\eqref{eq:subdef} for the leading-power soft function, 
but adapted to the case of an initial state gluon and anti-quark. 

We therefore define the \textit{subtracted} NLP soft operator and soft function as 
\begin{equation}\label{eq:uns/subop}
O_{g\bar{q}}^{\rm NLP}(x_0,\{s\}) \equiv \frac{O_{g\bar{q},\rm unsub}^{\rm NLP} (x_0, \{s\})}{S_{\sqcap}^{+}(x^0/2)\EuScript{S}_{\sqcap}^{-}(x^0/2)}\,, 
\end{equation}
\begin{equation}\label{eq:uns/sub}
   S_{g\bar{q}}^{\rm NLP} (x_0, \{s\}) \equiv \langle 0|O_{g\bar{q}}^{\rm NLP}(x_0,\{s\})|0 \rangle \,, 
\end{equation}
respectively, where $S_{\sqcap}^{+}(x^0/2)$ is given in~\eqref{eq:LP-subp} and 
$\EuScript{S}_{\sqcap}^{-}(t)$ is the vacuum matrix element of the generalization of \eqref{eq:Wdef} 
to the adjoint representation, 
\begin{equation}
	\label{eq:adjWdef}
	\EuScript{W}_{\sqcap}^{-}(t) \equiv \frac{1}{N_c^2-1}\, \mathbf{T}\left([\mathcal{Y}_{n_-}^\dagger]^{ab}(tn_+)\, [tn_+, 0]^{bc}\, \mathcal{Y}^{ca}_{n_-}(0)\,\right).
\end{equation}
By \eqref{eq:Wilsonlength}, we have
\begin{equation}
	\label{eq:finiteadj}
	\begin{aligned}
		[tn_+, 0]^{bc} = \widehat{\rm P}\exp\left[g_sf^{dbc}\int^t_0\df \lambda\, n_+\cdot A^d(\lambda n_+)\right].
	\end{aligned}
\end{equation} 
The calculation of the adjoint subtraction  gives 
\begin{align}
	\label{eq:adjsub}
	\EuScript{S}_{\sqcap}^{-}(t) &=\langle 0 | \EuScript{W}_{\sqcap}^{-}(t) | 0 \rangle = 1 + \frac{\alpha_s}{\pi}C_A \left[-\frac{1}{\eps^2}-\frac{1}{\eps}\left(\ln\frac{\mu}{\left|\delta_-\right|}+\ln\left(i\mu e^{\gamma_E}t\right)\right)\right]+\mathcal{O}(\alpha_s^2,\eps^0)\,,
\end{align}
and agrees with the one for the fundamental representation in \eqref{eq:LP-subm} up to the replacement $C_F\to C_A$ as expected. Combining \eqref{eq:NLPsoftuns}, 
\eqref{eq:LP-subp} and \eqref{eq:adjsub}, we find that the limits 
$\delta_\pm\to 0$ can be taken in $ S_{g\bar{q}}^{\rm NLP} (x_0, \{s\}) $ as defined in~\eqref{eq:uns/sub}, and hence the subtracted NLP soft operator has a well-defined UV anomalous dimension. 

%%%%%%%%%%%%%%%%%%%%%%%%%%%%%%%%%%%%%%%%%%%%%%%%%%%%
\begin{figure}[t]
\centering
\includegraphics[width=13cm]{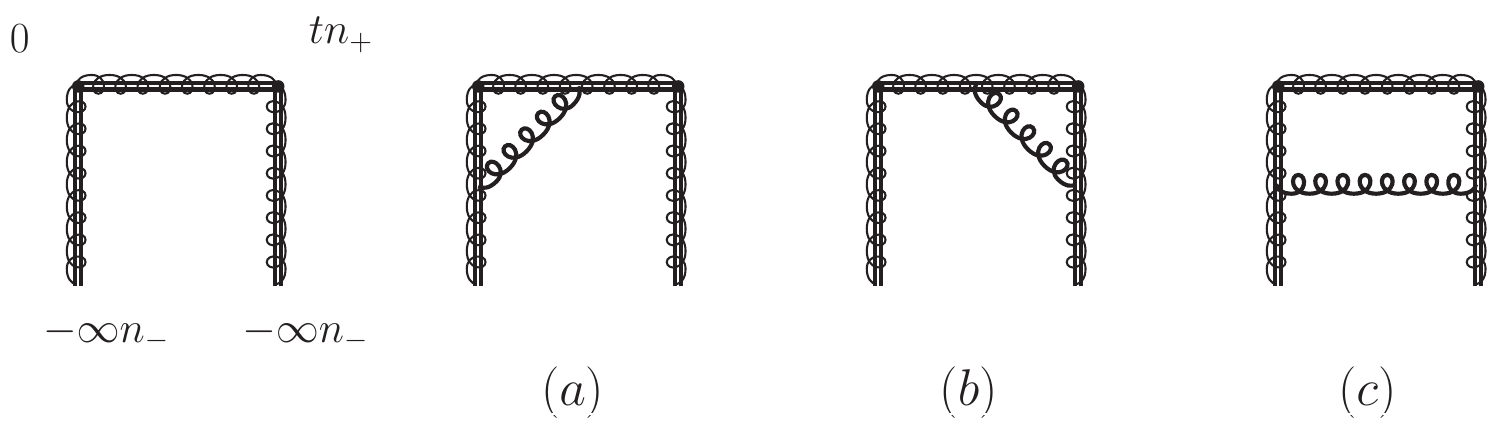}
\caption{\label{fig:NLP-sub}The adjoint soft subtraction operator and its one-loop corrections. Diagram $(c)$ does not contribute in the Feynman gauge.}
\end{figure}
%%%%%%%%%%%%%%%%%%%%%%%%%%%%%%%%%%%%%%%%%%%%%%%%%%%%%

The bare subtracted soft operator $O_{g\bar{q},\text{bare}}^{{\rm NLP}}(x_0,\{s\})$ is given by \eqref{eq:NLPsoftuns} divided by \eqref{eq:adjsub} with $t = x^0/2$ and $S_{\sqcap}^{+}(x^0/2)$. We then define the renormalized operator by 
\begin{align}
   \label{eq:NLPsoft}
	O_{g\bar{q}}^{\rm NLP}(x_0,\{s\};\mu) &= \left[Z_{g\bar{q}}^{\rm NLP}O_{g\bar{q},\text{bare}}^{{\rm NLP}}\right](x_0,\{s\}),
\end{align}
where the bare operator is defined as above in terms of the renormalized fields and  strong coupling, and the bracket emphasizes that the product inside is a convolution. Explicitly, the right-hand side reads
\begin{eqnarray}
   \label{eq:NLPsoftdetail}
	\left[Z_{g\bar{q}}^{\rm NLP}O_{g\bar{q},\text{bare}}^{{\rm NLP}}\right](x_0,\{s\}) &=& O_{g\bar{q},\text{bare}}^{{\rm NLP}}(x_0,\{s\}) -\frac{\alpha_s}{4\pi\eps} \Bigg\{-\beta_0+C_F\left[4L_{x_0}-2L_{s_1}-2L_{s_2}-1\right]\notag\\
	&& \hspace*{-3.3cm} + \,C_A\left[\frac{4}{\eps}+4L_{x_0}+L_{s_1}+L_{s_2}-\ln\frac{\partial_{s_1}\partial_{s_2}}{\mu^2}+2\right] \Bigg\}\,O_{g\bar{q},\text{bare}}^{{\rm NLP}}(x_0, \{s\})\\
	&&\hspace{-3.3cm}- \frac{\alpha_s}{2\pi\eps}\left(C_F-\frac{C_A}{2}\right)\int\limits_0^1\df u\left[\frac{u}{\bar{u}}\right]_+\Big[O_{g\bar{q},\text{bare}}^{{\rm NLP}}(x_0,u\,s_1, s_2)+O_{g\bar{q},\text{bare}}^{{\rm NLP}}(x_0,s_1, u\,s_2)\Big]+\mathcal{O}(\alpha_s^2),\notag
\end{eqnarray}
where $L_{x_0}$ is given by \eqref{eq:Lx0} and 
\begin{equation}
	\label{eq:Lz1Lz2}
	L_{s_1} = \ln\left(i\mu e^{\gamma_E} \left(s_1+i0^+\right)\right),\quad L_{s_2} = \ln\left(-i\mu e^{\gamma_E} \left(s_2-i0^+\right)\right).
\end{equation}
The $\pm i0^+$ prescriptions ensure the positivity of the corresponding  $\omega_1$ and $\omega_2$. 

When the soft-function computation is done employing  \eqref{eq:NLPsoftoperatordefnew} with $\mathcal{Y}^\dagger \to \mathcal{Y}^T$ 
in the anti-time ordered part, the subtraction operator \eqref{eq:adjWdef} should be changed similarly. Then in \eqref{eq:adjsub} one needs to replace $\ln\,(\mu/|\delta_-|)$ by $\frac{1}{2}\ln\,(\mu^2/(-\delta_-^2))$. This modification exactly compensates the corresponding one in the unsubtracted soft function.  Hence, irrespective of whether one starts with 
\eqref{eq:NLPsoftoperatordef1} or with
\eqref{eq:NLPsoftoperatordefnew} with $\mathcal{Y}^\dagger \to \mathcal{Y}^T$, the pole parts of the {\em subtracted} soft functions are identical and, consequently, the anomalous dimension is unique as it should be.

\subsection{Renormalization and RGE}
\label{sect:renorm}
In this section, we derive the anomalous dimension in different spaces based on \eqref{eq:NLPsoft}. In all cases, the anomalous dimensions in the $\overline{\rm MS}$ scheme can be derived from the renormalization factors according to
\begin{equation}
	\label{eq:NLPsoftAD}
	\gamma_{g\bar{q}}^{\rm NLP} \equiv - \left[\frac{\df\,Z_{g\bar{q}}^{\rm NLP}}{\df \ln\mu}\frac{1}{Z_{g\bar{q}}^{\rm NLP}}\right]\,,
\end{equation}
where the bracket implies a convolution of the two factors.

\subsubsection{Position space} 
\label{sect:posres}
The position-space renormalization factor in the $\overline{\rm MS}$ scheme is given explicitly in \eqref{eq:NLPsoft}, and 
the corresponding anomalous dimension/evolution kernel at the one-loop order is \begin{eqnarray}
    \left[\,\gamma^{\rm NLP}_{g\bar{q}}\,O_{g\bar{q}}^{\rm NLP}\,\right](x_0, \{s\};\mu) &=& -\frac{\alpha_s}{4\pi}\Bigg\{ \Bigg[8\left(C_F+C_A\right)L_{x_0}-4\left(C_F-\frac{C_A}{2}\right) \left(L_{s_1}+L_{s_2}\right)\notag\\
   &&\hspace*{-3cm}-\,2C_A\ln\frac{\partial_{s_1}\partial_{s_2}}{\mu^2}+4C_A-2C_F-2\beta_0\Bigg]O_{g\bar{q}}^{\rm NLP}(x_0, \{s\};\mu)\nonumber\\
   &&\hspace*{-3cm}+\,4\left(C_F-\frac{C_A}{2}\right)\int\limits_0^1\df u\left[\frac{u}{\bar{u}}\right]_+\Big[O_{g\bar{q}}^{{\rm NLP}}(x_0,u\,s_1, s_2;\mu)+O_{g\bar{q}}^{{\rm NLP}}(x_0, s_1, u\,s_2;\mu)\Big]\Bigg\}\qquad
  \label{eq:NLPsoftADpos}
\end{eqnarray}
with $\bar{u}=1-u$. This holds at the operator level. Taking the vacuum matrix element, the RGE for the position-space soft function is given by
\begin{equation}
	\label{eq:NLPsoftRGEpos}
	\frac{\df}{\df\ln\mu}S_{g\bar{q}}^{\rm NLP}(x_0,\{s\};\mu) = - \left[\,\gamma^{\rm NLP}_{g\bar{q}}\,S_{g\bar{q}}^{\rm NLP}\,\right](x_0,\{s\};\mu)\,. 
\end{equation}

\subsubsection{Mixed space}
\label{sect:mixed}
In this case, we only perform the Fourier transformations with respect to $\{s\}$ while leaving $x^0$ untouched. In mixed space we observe that the $x^0$ part is LP-like and the corresponding anomalous dimension and the RGE are local in $x^0$ while the NLP part is the same as the case of $gg\to h$~\cite{Beneke:2024cpq}. 

The renormalization factor in mixed-space can be obtained similarly from \eqref{eq:NLPsoftdetail} and formulae in~\cite{Beneke:2024cpq} as 
\begin{align}
   \label{eq:ZSmixed}
   Z_{g\bar{q}}^{\rm NLP}\left(x_0,\{\omega\},\{\omega'\}\right) &= \delta\left(\{\omega\}-\{\omega'\}\right)\notag\\[0.1cm]
   & \hspace{-2.8cm}-\frac{\alpha_s}{4\pi\eps}\Bigg\{\delta\left(\{\omega\}-\{\omega'\}\right)\bigg[\frac{2C_A}{\eps}+2 (C_F+C_A)L_{x_0}+2(C_F-C_A)\ln\frac{\omega_1}{\mu}+\frac{3C_F-\beta_0}{2}
  \bigg]\qquad\notag\\
   & \hspace{-1.35cm}+2\left(C_F-\frac{C_A}{2}\right)\delta(\omega_1-\omega_1')\omega_2\,\Gamma(\omega_2,\omega_2')+\Big(\omega_1\leftrightarrow\omega_2, \omega_1'\leftrightarrow\omega_2'\Big)\Bigg\}\,.
\end{align}
The anomalous dimension in the $\overline{\rm MS}$ scheme then follows as %{\color{blue}\bf[YJ: checked]}
\begin{align}
   \label{eq:NLPsoftADmixed}
   \gamma_{g\bar{q}}^{\rm NLP}(x_0,\{\omega\},\{\omega'\}) = &-\frac{\alpha_s}{4\pi}\Bigg\{\delta\left(\{\omega\}-\{\omega'\}\right)\left[4(C_F+C_A)L_{x_0}+4(C_F-C_A)\ln\frac{\omega_1}{\mu}+3C_F-\beta_0\right]\notag\\
  &\hspace{-2.2cm}+4\left(C_F-\frac{C_A}{2}\right)\delta(\omega_1-\omega_1')\omega_2\,\Gamma(\omega_2,\omega_2') \Bigg\}+\Big(\omega_1\leftrightarrow\omega_2, \omega_1'\leftrightarrow\omega_2'\Big)\,,
\end{align}
where the non-local kernel $\Gamma(\omega,\omega')$ is 
defined by the distribution
\begin{equation}
\label{eq:BLCDAkerneldef}
	\Gamma(\omega,\omega') = \underbrace{\left[\frac{\theta(\omega-\omega')}{\omega\left(\omega-\omega'\right)}\right]_+}_{\Gamma_{<}(\omega,\omega')}+ \underbrace{\left[\frac{\theta(\omega'-\omega)}{\omega'\left(\omega'-\omega\right)}\right]_+}_{\Gamma_{>}(\omega,\omega')}.
\end{equation}
We note that the dependence on $x^0$ is local in $\omega_{1,2}$. 
The RGE in this mixed-space reads
\begin{equation}
    \label{eq:RGEmixed}
    \frac{\df}{\df\ln\mu}S_{g\bar{q}}^{\rm NLP}(x_0, \{\omega\};\mu) = - \int\limits_0^\infty\df\omega_1'\int\limits_0^\infty\df\omega_2'\, \gamma_{g\bar{q}}^{\rm NLP}(x_0,\{\omega\},\{\omega'\}) S_{g\bar{q}}^{\rm NLP}(x_0, \{\omega'\};\mu)\,.
\end{equation}

\subsubsection{Momentum space}
\label{sect:momres}

The renormalized soft function in momentum space follows by Fourier-transforming the mixed-space results with respect to $x^0$. The relation between the renormalized and bare soft function is
\begin{align}
   \label{eq:NLPsoftmomrendef}
   S^{\rm NLP}_{g\bar{q}}(\Omega,\{\omega\};\mu) = \int\limits_0^\infty\df \Omega'\int\limits_{0}^{\infty}\{\df\omega'\}\,Z_{g\bar{q}}^{\rm NLP}\left(\Omega,\{\omega\},\Omega',\{\omega'\}\right)S^{\rm NLP}_{g\bar{q},\rm bare}(\Omega',\{\omega'\})\,,
\end{align}
and with the help of  results from \cite{Beneke:2024cpq}  we obtain 
\begin{eqnarray}
   Z_{g\bar{q}}^{\rm NLP}	&=& \delta(\{\omega\}-\{\omega'\})\bigg[\overbrace{\delta(\Omega-\Omega')+\frac{\alpha_s}{\pi \eps}(C_F+C_A)\left(\delta(\Omega-\Omega')\ln\frac{\Omega}{\mu}+\Omega\Gamma_<(\Omega,\Omega')\right)}^{\text{LP-like}}\bigg]\notag\\
   &&-\,\frac{\alpha_s}{4\pi\eps}\delta(\Omega-\Omega')\Bigg\{\left[\frac{2C_A}{\eps}+\frac{3C_F-\beta_0}{2}+2(C_F-C_A)\ln\frac{\omega_1}{\mu}\right]\delta(\{\omega\}-\{\omega'\})\notag\\
   && +\,2\left(C_F-\frac{C_A}{2}\right)\delta(\omega_1-\omega_1')\omega_2\,\Gamma(\omega_2,\omega_2')+\Big(\omega_1\leftrightarrow\omega_2, \omega_1'\leftrightarrow\omega_2'\Big)\Bigg\}\,.
   \label{eq:NLOsoftmomZS}
\end{eqnarray}
Then the $\overline{\text{MS}}$ momentum-space anomalous dimension reads 
\begin{eqnarray}
			\gamma_{g\bar{q}}^{\rm NLP}\left(\Omega,\{\omega\},\Omega',\{\omega'\}\right) &=& \frac{\alpha_s}{4\pi}\Bigg\{4(C_F+C_A)\left(\delta(\Omega-\Omega')\ln\frac{\Omega}{\mu}+\Omega\,\Gamma_<(\Omega,\Omega')\right)\delta\left(\{\omega\}-\{\omega'\}\right)\notag\\
		&&\hspace{-2.5cm}-\delta(\Omega-\Omega')\delta(\omega_2-\omega_2')\Bigg[\left(4(C_F-C_A)\ln\frac{\omega_1}{\mu}+3C_F-\beta_0\right)\delta(\omega_1-\omega_1')\notag\\
		&&\hspace{-2.5cm}+4\left(C_F-\frac{C_A}{2}\right)\omega_1\,\Gamma(\omega_1,\omega_1')\Bigg]\Bigg\}+\Big(\omega_1\leftrightarrow\omega_2, \omega_1'\leftrightarrow\omega_2'\Big)\,.
\label{eq:ADmom}
\end{eqnarray}
The corresponding RGE is given by
\begin{equation}
	\label{eq:NLPsoftRGEmom}
	\frac{\df}{\df\ln\mu}S_{g\bar{q}}^{\rm NLP}(\Omega,\{\omega\};\mu) = -\int\limits_0^\infty\df\Omega'\int\limits_0^\infty\{\df\omega'\}\,\gamma_{g\bar{q}}^{\rm NLP}\left(\Omega,\{\omega\},\Omega',\{\omega'\}\right)\,S_{g\bar{q}}^{\rm NLP}(\Omega',\{\omega'\};\mu)\,.
\end{equation} 

The first line in \eqref{eq:ADmom} is a LP-like contribution. We reproduce the LP soft anomalous dimension by the replacement $C_A\to C_F$ for that part. This is reasonable because, in the current case, two out of the four fundamental semi-infinite Wilson lines are replaced with adjoint ones. The second and third lines resemble the anomalous dimension of the $gg\to h$ soft function~\cite{Liu:2022ajh, Beneke:2024cpq}, which reads
\begin{align}
    \label{eq:ADgg2h1L}
	\gamma_g(w,w') &  = -\frac{\alpha_s}{4\pi}\left[\left(4(C_F-C_A)\ln\frac{w}{\mu^2}+6C_F\right)\delta(w-w')+8\left(C_F-\frac{C_A}{2}\right)w\,\Gamma(w,w')\right].
\end{align}
In App.~\ref{app:Bt3-to-hgg} we further establish a relation between 
the kernel for the three-particle twist-3 B-meson LCDA~\cite{Braun:2015pha} in the soft limit 
and the above kernels $\gamma_{g\bar{q}}^{\rm NLP}\left(\Omega,\{\omega\},\Omega',\{\omega'\}\right)$ and $\gamma_g(w,w')$, 
which is related to the universal appearance of the soft-quark 
operator \eqref{eq:octetblock}.

%%%%%%%%%%%%%%%%%%%%%%%%%%%%%%%%%%%%%%%%%%%%%%%%%%%%%%%%%%%%%%%%%

\section{Endpoint-finiteness of one-loop evolution}
\label{sect:endpoint}

Before proceeding to solving the RGE, we make an interesting 
observation. When the soft function is convoluted with the hard-collinear functions 
in the DY process, the convolution generates the endpoint divergences noted in \cite{Beneke:2019oqx,Broggio:2023pbu}. However, the convolution with the 
renormalization kernel remains finite. A similar protection of the evolution from endpoint divergences has been found in \cite{Beneke:2022obx} for the convolution of the hard function for next-to-leading-power jet production with the hard-collinear jet 
function. In this case as well, the convolution of the singular hard function with its anomalous dimension is well-defined, but not the convolution with the jet function 
itself.

To demonstrate the endpoint divergence in the convolution between the 
renormalized NLP soft-quark function and the jet functions in the collinear and anti-collinear sectors, see \eqref{eq:factnlpori}, it suffices 
to consider the leading-order expression \eqref{eq:SgqLOres} in four dimensions which reads
\begin{align}
	\label{eq:NLPsoftfuncMSbarLO}
	S^{\rm NLP}_{g\bar{q}}\left(\Omega,\{\omega\}; \mu\right) = \frac{\alpha_s}{4\pi}\,\omega_1\delta(\omega_1-\omega_2)\theta(\Omega-\omega_1). 
\end{align}
The jet functions at tree-level are proportional to $1/\omega$, so the 
convolution integral is 
\begin{equation}
   \frac{\alpha_s}{4\pi} \int\limits_0^\infty\df \omega_1 \int\limits_0^\infty\df \omega_2 \frac{1}{\omega_1}\,\frac{1}{\omega_2}\,\omega_1\delta(\omega_1-\omega_2)\theta(\Omega-\omega_1) =\frac{\alpha_s}{4\pi} \int\limits_0^\Omega \frac{\df\omega}{\omega} = \infty. 
\end{equation}
The endpoint divergence from the lower integration limit $\omega= 0$ can be subtracted by a cut-off as 
indicated in \eqref{eq:factnlpori}, which emerges naturally from rearrangements in the factorization formula. We do not elaborate here on the necessary rearrangements, but rather show that the convolution of the jet function with the renormalization kernel of the soft function is better behaved, hence the rearrangement commutes with 
renormalization-group evolution.

For this purpose, we use that the jet function to any order in perturbation 
theory depends on $\omega$ only in the form 
$1/\omega\times \ln^n(\omega/\mu)$, and investigate 
whether
\begin{equation}
    \label{eq:factandrgecont}
    \int\limits_0^\infty\frac{\df\omega_1}{\omega_1}\ln^n\frac{\omega_1}{\mu}\int\limits_0^\infty\frac{\df\omega_2}{\omega_2}\ln^m\frac{\omega_2}{\mu}\,\,\gamma_{g\bar{q}}^{\rm NLP}(\Omega,\{\omega\};\Omega',\{\omega'\})
\end{equation}
is finite. The terms in \eqref{eq:ADmom} involving delta functions in 
both $\omega_1$, $\omega_2$ are trivially finite. The non-trivial part is the remaining contribution from the third line of \eqref{eq:ADmom}, which reads
\begin{eqnarray}
    &&\int\limits_0^\infty\frac{\df\omega_1}{\omega_1}\ln^n\frac{\omega_1}{\mu}\int\limits_0^\infty\frac{\df\omega_2}{\omega_2}\ln^m\frac{\omega_2}{\mu}\,\delta(\omega_2-\omega_2')\omega_1 \Gamma(\omega_1,\omega_1')\nonumber\\
    &&=\, \frac{1}{\omega_2'}\ln^m\left(\frac{\omega_2'}{\mu}\right)\int\limits_0^\infty\df\omega_1\ln^n\left(\frac{\omega_1}{\mu}\right)\, \Gamma(\omega_1,\omega_1')
\end{eqnarray}
plus a similar integral with $\omega_1$ and $\omega_2$ exchanged.
The remaining integration can be performed with the help of
\begin{align}
    \int\limits_0^\infty\df \omega\,\omega^\eta \,\Gamma(\omega,\omega') =& \int\limits_0^\infty\df \omega\,(\omega^\eta-\omega'^{\,\eta})\left(\frac{\theta(\omega-\omega')}{\omega(\omega-\omega')}+\frac{\theta(\omega'-\omega)}{\omega'(\omega'-\omega)}\right)=\omega'^{-1+\eta}\left[-H_{-\eta}-H_{\eta}\right],
\end{align}
where $H_x$ is the harmonic-number function. Then
\begin{align}
    \int\limits_0^\infty\df \omega\,\ln^k\left(\frac{\omega}{\mu}\right)\,\Gamma(\omega,\omega') = \frac{1}{\omega'}\frac{\df^k}{\df \eta^k}\left[\left(\frac{\omega'}{\mu}\right)^{\!\!\eta}(-H_{\eta}-H_{-\eta})\right]_{\eta=0},
\end{align}
which is finite as was to be shown. 

%%%%%%%%%%%%%%%%%%%%%%%%%%%%%%%%%%%%%%%%%%%%%%%%%%%%%%%%%%%%%%%%%%%%%%%

\section{Renormalization-group improved soft function}
\label{sect:RGEsol}

Given the anomalous dimensions, one can solve the corresponding RGEs to resum large logarithms in the soft function when $\mu \gg \Omega,\omega_n$.

\subsection{Solution of the RGE in mixed space}

We first derive the solution in mixed space, since in this case the $x^0$-dependence is multiplicative. The anomalous dimension \eqref{eq:NLPsoftADmixed} involves the  distribution $\omega \,\Gamma(\omega,\omega')$. Its eigenfunctions are 
the power functions $\omega^a$, since
\begin{equation}
    \label{eq:eigenGamma}
    \int\limits_0^{\infty} \df\omega^{\prime}\, \omega\,\Gamma\left(\omega,\omega^{\prime}\right)\left({\omega^{\prime}}\right)^a=\underbrace{-\left[H_a+H_{-a}\right]}_{\equiv \mathcal{F}(a)}\omega^a=-\left[\psi(1+a)+\psi(1-a)+2\gamma_E\right]\omega^a\,.
\end{equation}
Here $H_x$ and $\psi(x)$ are the harmonic number and digamma function, respectively. We denote the eigenvalue as $\mathcal{F}(a)$. Hence the easiest way to solve the integro-RGE is to decompose the solution in terms of power functions of $\omega_1$ and $\omega_2$, which turns out to be the Mellin inversion with respect to $\eta_1$ and $\eta_2$ as follows:
\begin{equation}
    \label{eq:Mellininversion}
    \begin{aligned}
        S_{g\bar{q}}^{\rm NLP}(x_0,\{\omega\};\mu) &= \int\limits_{c_1-i\infty}^{c_1+i\infty}\frac{\df\eta_1}{2\pi i}\int\limits_{c_2-i\infty}^{c_2+i\infty}\frac{\df\eta_2}{2\pi i}\left(ie^{\gamma_E}\frac{x^0}{2}\omega_1\right)^{\!\!\!\!-\eta_1}\left(ie^{\gamma_E}\frac{x^0}{2}\omega_2\right)^{\!\!\!\!-\eta_2}\,S_{g\bar{q}}^{\rm NLP}(x_0,\{\eta\};\mu)\,.
    \end{aligned}
\end{equation}
As before, the $i0^+$ prescription on $x^0$ is suppressed for brevity. The factor of $(ie^{\gamma_E}x^0/2)^{-\eta_1-\eta_2}$ in the integrand is the canonical choice to preserve 
the dimension of the soft function. 
 The specific values of $c_1,c_2\in \mathbb{R}$ are placeholders for  integration contours\footnote{Depending on the order of integrations, $c_1$ ($c_2$) can be a function of $\eta_2$ ($\eta_1$).} determined by the analyticity and convergence properties of $S_{g\bar{q}}^{\rm NLP}(x_0,\{\eta\};\mu)$ in the complex plane, and follow the general prescriptions for the Mellin integrals: The integration path must appropriately separate the simple poles extending to the left from those extending to the right in the complex plane of $\eta_{1,2}$, not cross any branch cuts, and is allowed to deviate from a straight line to satisfy these requirements. In practice, the simple poles are commonly generated by the $\Gamma$ functions making the separation of poles simpler. More details regarding the integration path for the DY soft function will be provided in Sec.~\ref{sec:numerics}. 
The corresponding Mellin transformation from 
momentum space reads
\begin{equation}
    \label{eq:Mellintrans}
    \begin{aligned}
        S_{g\bar{q}}^{\rm NLP}(x_0,\{\eta\};\mu) &= \int\limits_0^\infty\frac{\df\omega_1}{\omega_1}\int\limits_0^\infty\frac{\df\omega_2}{\omega_2}\,S_{g\bar{q}}^{\rm NLP}(x_0,\{\omega\};\mu)\left(ie^{\gamma_E}\frac{x^0}{2}\omega_1\right)^{\eta_1}\left(ie^{\gamma_E}\frac{x^0}{2}\omega_2\right)^{\eta_2}\,.
    \end{aligned}
\end{equation}

To obtain the RGE in Mellin space, we begin from 
\begin{eqnarray}
    \frac{\rm d}{{\rm d}\ln\mu} S_{g\bar{q}}^{\rm NLP}(x_0,\{\eta\};\mu) &=& -\int\limits_0^\infty\{\df\omega\}\,\Bigg[ \int\limits_0^\infty\left\{\frac{\df\omega'}{\omega'}\right\}\, \gamma_{g\bar{q}}^{\rm NLP}(x_0,\{\omega'\},\{\omega\})\left(ie^{\gamma_E}\frac{x^0}{2}\omega'_1\right)^{\eta_1}\notag \\
   && \times \left(ie^{\gamma_E}\frac{x^0}{2}\omega'_2\right)^{\eta_2}\Bigg] S_{g\bar{q}}^{\rm NLP}(x_0, \{\omega\};\mu),
\end{eqnarray}
where $\{{\rm d\omega'}/\omega'\}=\df\omega'_1\df\omega'_2/(\omega'_1\omega'_2)$. The two-fold integration inside the square bracket can be performed to give
\begin{align}
\label{eq:mellincalc}
    &-\frac{\alpha_s}{4\pi}\frac{1}{\omega_1\omega_2}\left(ie^{\gamma_E}\frac{x^0}{2}\omega_1\right)^{\eta_1}\left(ie^{\gamma_E}\frac{x^0}{2}\omega_2\right)^{\eta_2}\bigg\{ 8(C_F+C_A)L_{x_0}+4(C_F-C_A)\ln\frac{\omega_1\omega_2}{\mu^2}\notag\\
    &\hspace{2cm}+2(3C_F-\beta_0)+4\left(C_F-\frac{C_A}{2}\right)\big[\mathcal{F}(\eta_1)+\mathcal{F}(\eta_2)\big]\bigg\}\\
    &= -\frac{\alpha_s}{4\pi}\frac{1}{\omega_1\omega_2}\bigg\{ 16C_A L_{x_0}+4(C_F-C_A)\big[\partial_{\eta_1}+\partial_{\eta_2}\big]+4\left(C_F-\frac{C_A}{2}\right)\big[\mathcal{F}(\eta_1)+\mathcal{F}(\eta_2)\big] \notag\\
    &\hspace{2cm}+2(3C_F-\beta_0)\bigg\}\left(ie^{\gamma_E}\frac{x^0}{2}\omega_1\right)^{\eta_1}\left(ie^{\gamma_E}\frac{x^0}{2}\omega_2\right)^{\eta_2}\notag,
\end{align}
such that
\begin{eqnarray}
    &&\frac{\rm d}{{\rm d}\ln\mu} S_{g\bar{q}}^{\rm NLP}(x_0,\{\eta\};\mu) = -\frac{\alpha_s}{4\pi}\bigg\{-16C_AL_{x_0}-4(C_F-C_A)\big[\partial_{\eta_1}+\partial_{\eta_2}\big]-2(3C_F-\beta_0)\nonumber\\
    &&\hspace*{1cm}-\,4\left(C_F-\frac{C_A}{2}\right)\big[\mathcal{F}(\eta_1)+\mathcal{F}(\eta_2)\big]\bigg\}S_{g\bar{q}}^{\rm NLP}(x_0,\{\eta\};\mu)\,.
\end{eqnarray}
In passing from \eqref{eq:mellincalc} to this equation, the derivatives with respect to the Mellin variables generate 
$\ln(x^0\omega_n)$, which converts the coefficient of 
$L_{x_0}$ to the adjoint colour factor $C_A$. 

One important difference from the LP soft function is the presence of a transcendental function of the Mellin variables. Almost all factors in the above expression  are related to cusp anomalous dimensions in the fundamental and adjoint representations,
\begin{equation}
    \label{eq:cusp}
    \Gamma_{\rm cusp}^{F}(\alpha_s) = C_F\, \gamma_{\rm cusp}(\alpha_s),\qquad\Gamma_{\rm cusp}^{A}(\alpha_s) = C_A\, \gamma_{\rm cusp}(\alpha_s),
\end{equation}
where $\gamma_{\rm cusp}=\alpha_s/\pi+\cdots$. 
Introducing the non-cusp anomalous 
dimension
\begin{equation}
\label{eq:gamnc}
\gamma_{\rm nc}(\alpha_s) = 
(-6C_F)\frac{\alpha_s}{4\pi} +\mathcal{O}(\alpha_s^2)
\end{equation}
and the QCD beta-function (definition in App.~\ref{appendix:RG}),
the above RGE can be cast into the form
\begin{align}
    \label{eq:ADMellin}
    \frac{\rm d}{{\rm d}\ln\mu} S_{g\bar{q}}^{\rm NLP}(x_0,\{\eta\};\mu) = -\gamma_{g\bar{q}}^{\rm NLP}(x_0,\{\eta\})\,S_{g\bar{q}}^{\rm NLP}(x_0,\{\eta\};\mu),
\end{align}
with
\begin{align}
    \label{eq:ADMellinansatz}
    \gamma_{g\bar{q}}^{\rm NLP}(x_0,\{\eta\}) =& -4C_A\gamma_{\rm cusp}(\alpha_s) L_{x_0}+\gamma_{\rm nc}(\alpha_s)-\beta(\alpha_s)\\
   &\hspace{-1cm}-\sum_{n=1}^2\left[\left(C_F-C_A\right)\gamma_{\rm cusp}(\alpha_s)\partial_{\eta_n}+\left(C_F-\frac{C_A}{2}\right)\gamma_{\rm cusp}(\alpha_s)\mathcal{F}(\eta_n)\right] + \mathcal{O}(\alpha_s^2)\notag.
\end{align}
The above RGE no longer contains an integration.\footnote{The $\mathcal{O}(\alpha_s^2)$ symbol indicates that at the two-loop order additional $\eta$-dependent kernels are expected to appear, as is the case for the leading-twist $B$-LCDA  \cite{Braun:2019wyx}. } Instead, it involves derivatives in $\eta_1$ and $\eta_2$, which act as translations in Mellin space. Similar to \cite{Liu:2020eqe}, we can reorganize the above RGE into
\begin{align}
    \label{eq:RGEMellinMod}
    &\left[\frac{\rm d}{{\rm d}\ln\mu}-\left(C_F-C_A\right)\gamma_{\rm cusp}(\alpha_s)\left(\partial_{\eta_1}+\partial_{\eta_2}\right)\right] S_{g\bar{q}}^{\rm NLP}(x_0,\{\eta\};\mu) = \Bigg[4C_A\gamma_{\rm cusp}(\alpha_s)L_{x_0}-\gamma_{\rm nc}(\alpha_s)\notag\\
    &\hspace{0.5cm}+ \beta(\alpha_s) + \left(C_F-\frac{C_A}{2}\right)\gamma_{\rm cusp}(\alpha_s)\left(\mathcal{F}(\eta_1)+\mathcal{F}(\eta_2)\right)\Bigg] S_{g\bar{q}}^{\rm NLP}(x_0,\{\eta\};\mu)\,.
\end{align}
The left-hand side implies that the resummed soft function develops a shift in the Mellin-space variables during scale evolution. To make this shift explicit, we introduce the RG functions~\cite{Becher:2006mr,Becher:2007ty} 
\begin{align}
   \label{eq:RGfunc}
   S_{\Gamma^R}(\nu, \mu) = -\int\limits_{\alpha_s(\nu)}^{\alpha_s(\mu)}\df \alpha\,\frac{\Gamma_{\rm cusp}^R(\alpha)}{\beta(\alpha)}\int\limits_{\alpha_s(\nu)}^\alpha\frac{\df\alpha'}{\beta(\alpha')},\quad a_{\gamma}(\nu,\mu) = -\int\limits_{\alpha_s(\nu)}^{\alpha_s(\mu)}\df \alpha\,\frac{\gamma(\alpha)}{\beta(\alpha)},
\end{align}
which satisfy the evolution 
equations 
\begin{equation}
    \label{eq:RGfuncscaleevo}
    \frac{\df}{\df\ln\mu}S_{\Gamma^R}(\nu, \mu) = -\Gamma_{\rm cusp}^R\big(\alpha_s(\mu)\big)\ln\frac{\mu}{\nu},\qquad \frac{\df}{\df\ln\mu}a_\gamma(\nu, \mu) = -\gamma\big(\alpha_s(\mu)\big)\,,
\end{equation}
with $\Gamma_{\rm cusp}^R=C_R\gamma_{\rm cusp}$ and  $C_R=C_F$ or $C_A$. Here $\gamma$ can be any anomalous dimension function. 
The perturbative expansions of the RG functions are collected in App.~\ref{appendix:RG}.  

With these definitions, the solution to the RGE \eqref{eq:RGEMellinMod}, which gives the 
scale evolution of $S_{g\bar{q}}^{\rm NLP}(x_0,\{\eta\};\mu)$, can be expressed as
\begin{equation}
\label{eq:sol1}
   S_{g\bar{q}}^{\rm NLP}(x_0,\{\eta\};\mu) = U_S(x_0,\{\eta\};\mu_s,\mu)\,S_{g\bar{q}}^{\rm NLP}\Big(x_0,\{\eta-a_{\Gamma^F}(\mu_s,\mu)+a_{\Gamma^A}(\mu_s,\mu)\};\mu_s\Big)\,,\end{equation}
which makes the shift of the Mellin variables by the amount  $a_{\Gamma^F}-a_{\Gamma^A}$ manifest. Apart from this important point, the solution takes the same form as the standard Sudakov evolution. 
After inserting \eqref{eq:sol1} into \eqref{eq:RGEMellinMod}, we obtain for $U_S$ the same RGE as for the soft function itself, with the boundary condition 
\begin{equation}
    U_S(x_0,\{\eta\};\mu_s, \mu_s)=1\,,
\end{equation}
while the solution for $U_S$ reads
\begin{align}
    \label{eq:USsol}
    U_S(x_0,\{\eta\};\mu_s,\mu) &= r\,\exp\left[-4S_{\Gamma^{A}}(\mu_s,\mu)+a_{\gamma_{\rm nc}}(\mu_s,\mu)\right]\left(i\mu_s e^{\gamma_E}\frac{x^0}{2}\right)^{\!\!\!\!-4a_{\Gamma^{A}}(\mu_s,\mu)}\\
    & \hspace{-2cm}\times\exp\Bigg[\,\,\int\limits_{\alpha_s(\mu_s)}^{\alpha_s(\mu)}\frac{\rm d\alpha}{\beta(\alpha)}\left(\Gamma_{\rm cusp}^{F}(\alpha)-\frac{1}{2}\Gamma_{\rm cusp}^{A}(\alpha)\right)\sum_{n=1}^2\mathcal{F}\Big(\eta_n-a_{\Gamma^{F}}(\mu_\alpha,\mu)+a_{\Gamma^{A}}(\mu_\alpha,\mu)\Big)\Bigg]\notag
\end{align}
where $\mu_\alpha$ is defined as $\alpha_s(\mu_\alpha)=\alpha$ with $\alpha$ the variable of integration and
\begin{align}
    r=\frac{\alpha_s(\mu)}{\alpha_s(\mu_s)}\,.
\end{align}
The shifts inside $\mathcal{F}$ are again due to the derivatives with respect to $\eta_n$. 

To solve the RGE at LO in renormalization-group-improved perturbation theory, the 
LO soft function at the initial scale $\mu_s$ must be evolved with the evolution factors 
computed with the full anomalous-dimension kernel at the one-loop order and the 
two-loop cusp anomalous dimensions in the evaluation of $S_\Gamma$. The LO soft 
function in Mellin space is obtained from $S_{g\bar{q}}^{\rm NLP}(x_0,\{\omega\};\mu_s)$ in \eqref{eq:NLPsoftfuncMSbarLO} and the definition \eqref{eq:Mellintrans} as
\begin{equation}
    \label{eq:SMellinLO}
    S_{g\bar{q}}^{\rm NLP}(x_0,\{\eta\};\mu_s) = \frac{\alpha_s(\mu_s)}{4\pi}\left(i\frac{x^0}{2}\right)^{\!\!\!\!-1}\Gamma(\eta_1+\eta_2)\,e^{(\eta_1+\eta_2)\gamma_E}.
\end{equation}
Then we use \eqref{eq:Mellininversion} and shift the Mellin variables by $-a_\Gamma^-$ (defined below) to obtain the resummed soft function $S_{g\bar{q}}^{\rm NLP}(x_0,\{\omega\};\mu)$ in the form
\begin{eqnarray}
    S_{g\bar{q}}^{\rm NLP}(x_0,\{\omega\};\mu) 
    &=& V_S(x_0, \{\omega\};\mu_s,\mu)\left(i\frac{x^0}{2}\right)^{\!\!\!\!-1}\left[ \prod\limits_{n=1}^{2}\int\limits_{c_n+a_\Gamma^--i\infty}^{c_n+a_\Gamma^-+i\infty}\frac{\df\eta_n}{2\pi i}\left(i\frac{x^0}{2}\omega_n\right)^{\!\!\!\!-\eta_n}\right]\Gamma(\eta_1+\eta_2)\notag\\
    &&\hspace{-3cm}\times\exp\Bigg[\,\,\int\limits_{\alpha_s(\mu_s)}^{\alpha_s(\mu)}\frac{\rm d\alpha}{\beta(\alpha)}\left(\Gamma_{\rm cusp}^{F}(\alpha)-\frac{1}{2}\Gamma_{\rm cusp}^{A}(\alpha)\right)\sum_{n=1}^2\mathcal{F}\Big(\eta_n+a_{\Gamma^{F}}(\mu_s,\mu_\alpha)-a_{\Gamma^{A}}(\mu_s,\mu_\alpha)\Big)\Bigg],\;\qquad
    \label{eq:resummed1}
\end{eqnarray}
where \begin{eqnarray}
\label{eq:VS}
    V_S(x_0,\{\omega\};\mu_s,\mu) &=& \frac{\alpha_s(\mu)}{4\pi}\,\exp\big[-4S_{\Gamma^{A}}(\mu_s,\mu)+a_{\gamma_{\rm nc}}(\mu_s,\mu)\big]\left(\frac{\omega_1\omega_2}{\mu_s^2}\right)^{\!a^-_\Gamma}\left(i\mu_s e^{\gamma_E}\frac{x^0}{2}\right)^{\!-2a^+_{\Gamma}}\notag\\
    &=& \frac{\alpha_s(\mu)}{4\pi}\exp\left\{-\frac{4C_A}{\beta_0^2}\left[\frac{4\pi}{\alpha_s(\mu_s)}\left(1-\frac{1}{r}-\ln r\right)+\left(\frac{\Gamma_1^A}{\Gamma^A_0}-\frac{\beta_1}{\beta_0}\right)(1-r+\ln r)\right.\right.\notag\\
   &&\hspace{0cm}\left.\left.+\frac{\beta_1}{2\beta_0}\ln^2r\right]+\frac{\gamma_{\rm nc,0}}{2\beta_0}\ln r\right\}\left(\frac{\omega_1\omega_2}{\mu_s^2}\right)^{\!a_\Gamma^-}\left(i\mu_s e^{\gamma_E}\frac{x^0}{2}\right)^{\!{-}2a_\Gamma^+},
\end{eqnarray}
and
\begin{align}
    \rho_\Gamma &=\frac{C_A-2C_F}{2(C_A-C_F)}=\frac{1}{N_c^2+1}\stackrel{\rm QCD}{=}\frac{1}{10}\,,\\[0.1cm]
    a^\pm_{\Gamma} &= a_{\Gamma^A}(\mu_s,\mu)\pm a_{\Gamma^F}(\mu_s,\mu)=\frac{2(C_A\pm C_F)}{\beta_0}\ln r +\mathcal{O}(\alpha_s)
\end{align}
as well as $\gamma_{\rm nc,0}=-6 C_F$ from \eqref{eq:gamnc}.
In the above, we used formulae from App.~\ref{appendix:RG} to perturbatively expand the RG functions but did not insert the explicit expressions for the two-loop cusp anomalous dimension coefficient $\Gamma_1^A$, and the QCD beta-function coefficients $\beta_0$ and $\beta_1$ for brevity. The exponential in the last line of \eqref{eq:resummed1} 
can be simplified by changing variables from $\alpha$ to $a \equiv a_{\Gamma^{F}}(\mu_s,\mu_\alpha)-a_{\Gamma^{A}}(\mu_s,\mu_\alpha)$ and using 
\cite{Liu:2021mac,Liu:2022ajh}
\begin{align}
    &\,\exp\Bigg[\,\,\int\limits_{\alpha_s(\mu_s)}^{\alpha_s(\mu)}\frac{\rm d\alpha}{\beta(\alpha)}\left(\Gamma_{\rm cusp}^{F}(\alpha)-\frac{1}{2}\Gamma_{\rm cusp}^{A}(\alpha)\right)\sum_{n=1}^2\mathcal{F}\Big(\eta_n+a_{\Gamma^{F}}(\mu_s,\mu_\alpha)-a_{\Gamma^{A}}(\mu_s,\mu_\alpha)\Big)\Bigg]\notag\\
     &= \, \exp\Bigg[\,-\rho_\Gamma\sum_{n=1}^2\int\limits_0^{-a_\Gamma^-}{\rm d} a\,\mathcal{F}\Big(\eta_n+a\Big)+\mathcal{O}(\alpha_s)\Bigg]\,,
\end{align}
which holds to leading order in RG-resummed perturbation theory. The integral of the harmonic-number function contained in $\mathcal{F}$ defined in \eqref{eq:eigenGamma} is 
\begin{align}
    \exp\left(\int\df{x}\, H_x\right) = \exp\left(\ln\Gamma(1+x)-\gamma_E x\right),
\end{align} 
where ${\rm ln\Gamma}(x)$ is the so-called log-gamma function, which is identical to $\ln(\Gamma(x))$ for $x>0$.\footnote{However, while the former is an analytic function in the entire complex plane except for a single branch cut along the negative real axis, the latter possesses a more complex branch cut structure.}
We finally arrive at 
\begin{eqnarray}
    S_{g\bar{q}}^{\rm NLP}(x_0,\{\omega\};\mu) 
    &=&  \,V_S(x_0, \{\omega\};\mu_s,\mu)\left(i\frac{x^0}{2}\right)^{\!\!\!\!-1}\Bigg[\, \prod\limits_{n=1}^{2}\int\limits_{c_n+a_\Gamma^--i\infty}^{c_n+a_\Gamma^-+i\infty}\frac{\df\eta_n}{2\pi i}\left(i\frac{x^0}{2}\omega_n\right)^{\!\!\!\!-\eta_n}\Bigg]\Gamma(\eta_1+\eta_2)\notag \\
    &&\hspace{-2cm}\times \prod\limits_{n=1}^{2}\left[\frac{\Gamma(1+\eta_n-a_\Gamma^-)\Gamma(1-\eta_n)}{\Gamma(1-\eta_n+a_\Gamma^-)\Gamma(1+\eta_n)}e^{-2\gamma_E a_\Gamma^-}\right]^{\rho_\Gamma},
    \label{eq:resummed}
\end{eqnarray}
where the gamma-function product in the second line should be understood as 
\begin{align}
\left[\frac{\Gamma(x_1)\cdots\Gamma(x_n)}{\Gamma(y_1)\cdots\Gamma(y_n)}\right]^p = 
\exp\left\{p\left[\left({\rm ln\Gamma}(x_1)+\cdots {\rm ln\Gamma}(x_n)\right)-\left({\rm ln\Gamma}(y_1) + {\rm ln\Gamma}(y_n)\right)\right]\right\}.
\end{align}

Before investigating the effect of evolution numerically in Sec.~\ref{sec:numerics}, 
we conclude this section with a brief discussion of the solution in the large-$N_c$ and the QED ($C_A\to 0$) limits. 

\subsubsection{Large-$N_c$ limit}\label{sec:largeNc}

In the large-$N_c$ limit, $C_F=C_A/2$ and 
$\rho_\Gamma=0$, ant the Mellin inverse \eqref{eq:resummed} simplifies considerably, since the last factor reduces to 1. We find\footnote{In $V_S(x_0;\{\omega\};\mu_s,\mu)$, 
the appearing constants should also be evaluated in the large-$N_c$ limit, but the 
form of $V_S$ does not change.}
\begin{equation}
    S_{g\bar{q}}^{\rm NLP}(x_0,\{\omega\};\mu) 
    = V_S(x_0,\{\omega\};\mu_s,\mu)\left(i\frac{x^0}{2}\right)^{\!\!\!\!-1}\Bigg[\, \prod\limits_{n=1}^{2}\int\limits_{c_n+a_\Gamma^--i\infty}^{c_n+a_\Gamma^-+i\infty}\frac{\df\eta_n}{2\pi i}\left(i\frac{x^0}{2}\omega_n\right)^{\!\!\!\!-\eta_n}\Bigg]\Gamma(\eta_1+\eta_2).
\end{equation}
Changing variable $\eta_2\to \eta_2+\eta_1$, the $\eta_2$ integral becomes the inverse Mellin transform of the gamma function, and yields
\begin{eqnarray}
    \label{eq:resummedmasterformulaNc}
    S_{g\bar{q}}^{\rm NLP}(x_0,\{\omega\};\mu) 
    &=& V_S(x_0,\{\omega\};\mu_s,\mu)\left(i\frac{x^0}{2}\right)^{\!\!\!\!-1}\int\limits_{c_1+a_\Gamma^--i\infty}^{c_1+a_\Gamma^-+i\infty}\frac{\df\eta_1}{2\pi i}\left(\frac{\omega_1}{\omega_2}\right)^{\!\!\!\!-\eta_1}\,e^{-i\omega_2x^0/2}\notag\\
    &=& V_S(x_0,\{\omega\};\mu_s,\mu)\left(i\frac{x^0}{2}\right)^{\!\!\!\!-1}\delta\left(\omega_1-\omega_2\right)\omega_2\, e^{-i\omega_2 x^0/2}\,.
\end{eqnarray}
When $\mu=\mu_s$, $V_S(x_0,\{\omega\};\mu_s,\mu_s)=\alpha_s(\mu_s)/(4\pi)$, and we recover the LO result in \eqref{eq:SgqLOpartialF}  as required. The above result is a direct consequence of the large-$N_c$ limit of the anomalous dimension  \eqref{eq:NLPsoftADmixed}, where the non-local kernel proportional to $\Gamma(\omega_n,\omega_n')$ vanishes. Hence, in mixed space the scale evolution is local.  

\subsubsection{QED limit} 

In the QED limit $\rho_\Gamma\big|_{C_A=0} = 1$, 
and the second line of \eqref{eq:resummed} does not develop a branch cut. This is 
related to the fact that in this limit the one-loop anomalous dimension \eqref{eq:ADmom} reduces to 
\begin{eqnarray}
\label{eq:ADmomQED}
&& \frac{\alpha_s}{4\pi}\,\Bigg\{-\delta(\Omega-\Omega')\delta(\omega_2-\omega_2')\Bigg[\left(4C_F\ln\frac{\omega_1}{\mu}+3C_F-\beta_0\right)\delta(\omega_1-\omega_1')+4C_F\omega_1\,\Gamma(\omega_1,\omega_1')\Bigg]\quad\nonumber\\
 && \hspace*{0.1cm}+\,4C_F\left(\delta(\Omega-\Omega')\ln\frac{\Omega}{\mu}+\Omega\,\Gamma_<(\Omega,\Omega')\right)\delta\left(\{\omega\}-\{\omega'\}\right)
		\Bigg\}+\Big(\omega_1\leftrightarrow\omega_2, \omega_1'\leftrightarrow\omega_2'\Big)\,,
\end{eqnarray}
where now the coefficients of $\ln(\omega_1/\mu)$ and $\omega_1\Gamma(\omega_1,\omega_1')$ share the same cusp anomalous dimension coefficient in the fundamental representation. This 
degeneracy is exactly the same as in the transition from the anomalous dimension of the $gg\to h$ process (induced by light-quark loops) to that of $\gamma\gamma\to h$ \cite{Liu:2022ajh,Beneke:2024cpq}. 
The local evolution factor~\eqref{eq:VS} simplifies considerably due to 
\begin{equation}
    a_\Gamma^{\pm}\to \pm a_\Gamma^{\rm QED} \equiv \pm\frac{2C_F}{\beta_0}\ln r.
\end{equation}
and a cancellation of the double-logarithmic terms in the exponent, leaving only 
\begin{align}
   \label{eq:VSQED}
   V_S(x_0,\{\omega\};\mu_s,\mu) &\to \frac{\alpha_s(\mu)}{4\pi}r^{-\frac{3C_F}{\beta_0}}\left[\omega_1\omega_2 e^{2\gamma_E}\left(i\frac{x^0}{2}\right)^{\!2}\, \right]^{-a_\Gamma^{\rm QED
   }}\,.
\end{align}
In the non-local terms in \eqref{eq:SMellinLO}, the first Mellin integral results in the Meijer-G function, which can be reduced to some combinations of ${}_2F_2$ hypergeometric functions. However, we did not find a closed expression for the subsequent second Mellin integral. 

\subsubsection{UV poles of the soft function at $O(\alpha_s^2)$}

The DY NLP soft-quark function has been computed to next-to-leading order in 
$\alpha_s$ in \cite{Broggio:2023pbu} in dimensional regularization. The result 
is given before expansion in $\eps$, which would result in ultraviolet (UV) and infrared (IR) poles. To facilitate the future separation into UV and IR poles, 
we compute the UV poles of the soft function at $O(\alpha_s^2)$ by convoluting 
the renormalization factor \eqref{eq:NLOsoftmomZS} with the leading-order 
soft function \eqref{eq:SgqLOres} (here in $d$ dimensions),
\begin{equation}
S_{g\bar{q}}^{\rm NLP}(\Omega,\{\omega\};\mu)\big|_{O(\alpha_s^2), {\rm UV \,\,pole}} =-\Big[Z_{g\bar{q}}^{\rm NLP}S_{g\bar{q}}^{\rm NLP}(\Omega,\{\omega\};\mu)\big|_{O(\alpha_s)}\Big].
\end{equation}
The straightforward calculation results in 
\begin{eqnarray}
S_{g\bar{q}
}^{\rm NLP}(\Omega,\{\omega\};\mu)\big|_{O(\alpha_s^2), {\rm UV \,\,pole}} &=& 
\left(\frac{\alpha_s}{4\pi}\right)^{\!2} \omega_1\theta(\Omega-\omega_1)\nonumber\\[0.2cm]
&&\hspace*{-4cm}\times\,\Bigg\{
\delta(\omega_1-\omega_2)\Bigg[\,\frac{2 C_A}{\eps^2}+\frac{1}{\eps} 
\bigg(\!
-4 C_A\ln\frac{\omega_1(\Omega-\omega_1)}{\mu^2}-2 C_F\ln\frac{\Omega-\omega_1}{\omega_1}+\frac{3 C_F-\beta_0}{2}
\bigg)\Bigg]
\nonumber\\
&&\hspace*{-3.2cm}+\,\frac{2}{\eps}\left(C_F-\frac{C_A}{2}\right) \omega_2\,\Gamma(\omega_2,\omega_1)
\Bigg\}+\left(\omega_1\leftrightarrow\omega_2\right)\,.
\label{eq:UVpolesat2ndorder}
\end{eqnarray}

As mentioned earlier, Ref.~\cite{Broggio:2023pbu} employs a different definition of 
the soft function with $1/(i\nm\partial)$ acting on the dressed soft (anti-) quark fields 
in \eqref{eq:NLPsoftoperatordef1}. 
Accounting for this difference and a different colour normalization, the tree-level soft function in that convention is 
given by \eqref{eq:SgqLOres} divided by $2\omega_1\omega_2$. Since the relation by this factor is independent of the scale $\mu$, the renormalization kernel is not affected by the 
different definition. Hence, the UV poles of the soft function at $O(\alpha_s^2)$ as defined in \cite{Broggio:2023pbu} are given by multiplying the first term of \eqref{eq:UVpolesat2ndorder} by $1/(2\omega_1^2)$ and the second term with $\omega_1$ and $\omega_2$ exchanged accordingly by $1/(2\omega_2^2)$. 

\subsection{Numerical analysis}
\label{sec:numerics}

Factorization theorems are usually formulated in momentum space. Therefore, it is more desirable to have the evolved soft function fully in momentum space. 
To this end, we Fourier-transform \eqref{eq:resummed}  with respect to $x^0$, following the definition \eqref{eq:LPsoftmomdef}. Applying \eqref{eq:FourierOmega} we obtain
\begin{align}
    S_{g\bar{q}}^{\rm NLP}(\Omega,\{\omega\};\mu) = W_S(\mu_s,\mu) I(\Omega,\{\omega\}; \mu_s,\mu)
\end{align}
where
\begin{align}
W_S(\mu_s,\mu)  = \frac{\alpha_s(\mu)}{4\pi}\exp\bigg\{&-\frac{4C_A}{\beta_0^2}\left[\frac{4\pi}{\alpha_s(\mu_s)}\left(1-\frac{1}{r}-\ln r\right)+\left(\frac{\Gamma_1^A}{\Gamma^A_0}-\frac{\beta_1}{\beta_0}\right)(1-r+\ln r)\right.\notag\\
   &+\frac{\beta_1}{2\beta_0}\ln^2r\bigg]+\frac{\gamma_{\rm nc,0}}{2\beta_0}\ln r\bigg\}\,,
   \label{eq:solW}
\end{align}
and
\begin{eqnarray}
   I(\Omega,\{\omega\}; \mu_s,\mu) &=& \left(\frac{\omega_1\omega_2}{\mu_s^2}\right)^{\!a_\Gamma^-}\left(\frac{\Omega}{\mu_s}\right)^{\!2a_\Gamma^+}\left[ \prod\limits_{n=1}^{2}\int\limits_{c_n-i\infty}^{c_n+i\infty}\frac{\df\eta_n}{2\pi i}\left(\frac{\Omega}{\omega_n}\right)^{\!\!\eta_n}\right]
   \frac{\Gamma(\eta_1+\eta_2)\,e^{-2\gamma_E a_\Gamma^+}}{\Gamma(
   \eta_1+\eta_2+1+2a_\Gamma^+)}\notag\\
   &&\hspace*{0cm}\times \prod\limits_{n=1}^{2}
   \left[\frac{\Gamma(1+\eta_n-a_\Gamma^-)\Gamma(1-\eta_n)}{\Gamma(1-\eta_n+a_\Gamma^-)\Gamma(1+\eta_n)}\,e^{-2\gamma_E a_\Gamma^-}\right]^{\rho_\Gamma}.\quad
   \label{eq:I-int}
\end{eqnarray}
It is easy to check analytically that when $\mu=\mu_s$, we recover the leading-order initial momentum-space soft function~\eqref{eq:NLPsoftfuncMSbarLO}.\footnote{The numerical evaluation of the soft function at $\mu$ close to the initial scale $\mu_s$ suffers from uncertainties due to the $\delta(\omega_1-\omega_2)$ function at $\mu_s$. This issue can be resolved by writing 
$\displaystyle
     I(\Omega,\{\omega\}; \mu_s,\mu) = \frac{\df}{\df \omega_2}\int^{\omega_2}_0 \df\omega_2'\, I(\Omega,\omega_1, \omega_2'; \mu_s,\mu)
$, then solving directly for the integral and performing the derivative numerically. } 

\begin{figure}[t]
\centering
\includegraphics[width=0.6\textwidth]{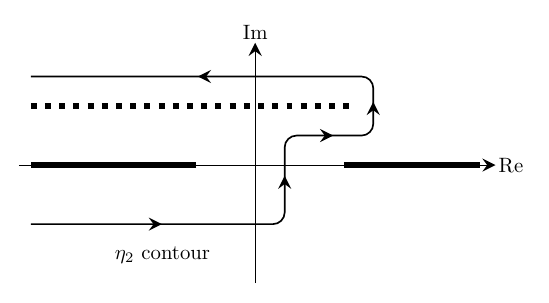}
\caption{\label{fig:integration-contour} Integration contour for $\eta_2$ at fixed $\eta_1$ when $\Omega/\omega_2>1$. The thick lines on the real axis represent the branch cuts (including branch points) of the integrand in~\eqref{eq:I-int}. The thick points represent the poles from the $\Gamma(\eta_1+\eta_2)$ factor.}
\end{figure}

Next, we have to fix the integration contours. Since the integrand decays slowly, or diverges for $\eta_n\to \pm i \infty$, the contour must be a closed loop beginning and ending at $+\infty$ or $-\infty$ while separating the singularities from the gamma functions in the numerator. The choice here depends on the value of the ratio $\Omega/\omega_n$ entering the integrand. In particular, the loop goes with a clockwise orientation around $+\infty$ for $\Omega/\omega_n < 1$ but counter-clockwise around $-\infty$ for $\Omega/\omega_n > 1$. Moreover, when evaluating the $\eta_2$ integral with a fixed $\eta_1$, the contour has to be chosen in such a way that the singularities arising from $\Gamma(\eta_1 + \eta_2)$ and $\Gamma(1 + \eta_2 - a_{\Gamma}^-)$ are on the other side of the contour than the singularities arising from $\Gamma(1 - \eta_2)$. One such contour is shown in Fig.~\ref{fig:integration-contour}.

%%%%%%%%%%%%%%%%%%%%%%%%%%%%%%%%%%%%%%%%%%%%%%%%%%%%%%%%%%%%%%%%%%%%%%
%%%%%%%%%%%%%%%%%%%%%%%%%%%%%%%%%%%%%%%%%%%%%%%%%%%%%%%%%%%%%%%%%%%%%% 

The numerical results shown in the following are obtained by employing 3-loop evolution of the strong coupling with $n_f=5$ active flavours and 
$\alpha_s(m_Z)=0.118$. We do not take the large-$N_c$ limit for evolving $\alpha_s(\mu)$, since this would entail $n_f=0$ and a significant departure from standard coupling values below $\mu=m_Z$. We fix the value of $\Omega$ to $10~{\rm GeV}$, and set the initial scale for the evolution of the tree-level soft function 
$S_{g\bar{q}}^{{\rm NLP}}(\Omega,\{\omega\}; \mu_s)$ to its natural scale  
$\mu_s=\Omega=10~{\rm GeV}$.

\subsubsection{Large-$N_c$ limit}

It is instructive to begin with a discussion of the large-$N_c$ limit. From \eqref{eq:solW} 
and \eqref{eq:I-int} or by Fourier-transforming \eqref{eq:resummedmasterformulaNc}, we obtain the evolved large-$N_c$ momentum-space soft function
\begin{eqnarray}
S_{g\bar{q},\,\rm large-{\it N}_c}^{\rm NLP}(\Omega,\{\omega\};\mu) &=& 
W_S(\mu_s,\mu)
\delta(\omega_1 - \omega_2)\theta(\Omega - \omega_2)\notag\\
&&\hspace{0cm}\times\,\omega_2\left(\frac{\omega_1\omega_2}{\mu_s^2}\right)^{\!a_\Gamma^-}\left(\frac{\Omega-\omega_{2}}{\mu_s}\right)^{\!2a_\Gamma^+}
\frac{e^{-2\gamma_E a_\Gamma^+}}{\Gamma(1+2a_{\Gamma}^+)} \,, 
\label{eq:largeNcmomentumspacesol}
\end{eqnarray}
where $W_S$ is understood as its large-$N_c$ limit, in which case $a_\Gamma^+ = \frac{3 C_A}{\beta_0} \ln\frac{\alpha_s(\mu)}{\alpha_s(\mu_s)}$.\footnote{Strictly speaking, \eqref{eq:largeNcmomentumspacesol} is valid only for $0 < 1 + 2a_{\Gamma}^+ < 1$, since the Fourier transform \eqref{eq:FourierOmega} holds only on this interval. Outside this range, singular terms arise at $\omega=\Omega$.}  

%%%%%%%%%%%%%%%%%%%%%%%%%%%%%%%%%%%%%%%%%%%%%%%%
\begin{figure}[t]
    \centering
    \includegraphics[width=0.45\textwidth]{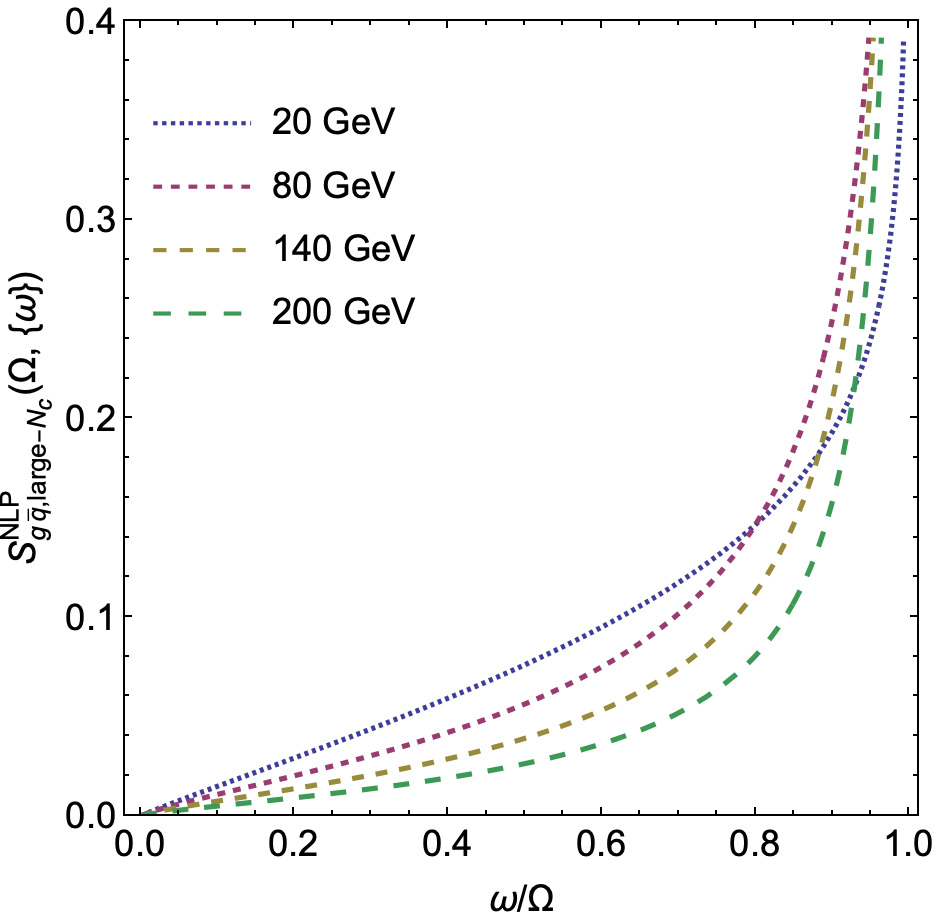}
    \hskip0.8cm
    \includegraphics[width=0.45\textwidth]{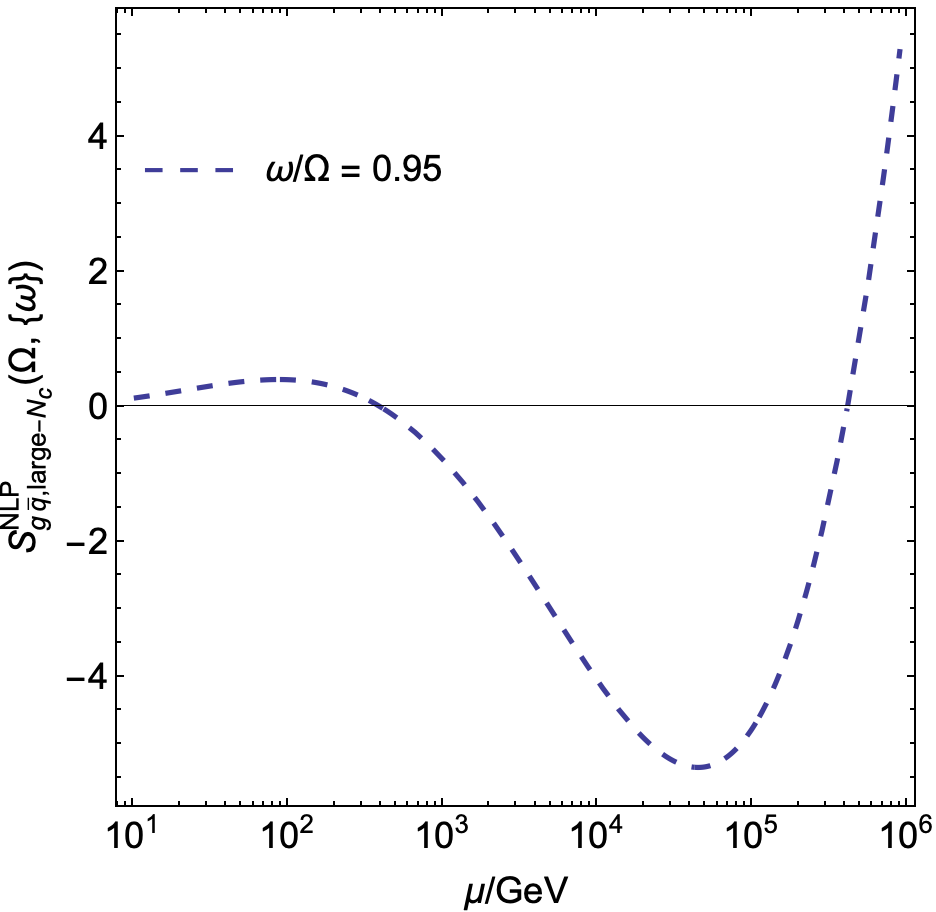}
   \caption{Shown is the evolution of the NLP soft function $S_{g\bar{q}}^{\rm NLP}(\Omega,\{\omega\};\mu)$ in the large $N_c$ limit with the delta function $\delta(\omega_1-\omega_2)$ stripped off. The left panel depicts the behaviour for different factorization scales along the diagonal where $\omega \equiv \omega_1 = \omega_2$. The right panel shows how the soft function evolves for the fixed value 0.95 of the ratio $\omega/\Omega$.}
    \label{fig:diagonal-section}
\end{figure}
%%%%%%%%%%%%%%%%%%%%%%%%%%%%%%%%%%%%%%%%%%%%%%%%%

The large-$N_c$ soft function is localized on the diagonal $\omega\equiv \omega_1=\omega_2$ with $0\leq\omega\leq \Omega$. Initially (at $\mu_s$) it grows linearly with $\omega$. 
When evolved from $\mu_s$ to a larger scale $\mu$, 
$a_\Gamma^+$ is negative, and the soft function at the scale $\mu$ develops a singularity 
at $\omega=\Omega$ as seen in \eqref{eq:largeNcmomentumspacesol} and the left panel of Fig. \ref{fig:diagonal-section}.
The origin of this singularity can be traced to the singular scale dependence of the (four-dimensional) tree-level soft function at the initial scale, which is given (for general $N_c$) by 
\begin{eqnarray}
\frac{\df}{\df\ln\mu}S_{g\bar{q}}^{\rm NLP}(\Omega,\{\omega\};\mu)\big|_{\mu = \mu_s} &=& -\int\limits_0^\infty\df\Omega'\int\limits_0^\infty\{\df\omega'\}\,\gamma_{g\bar{q}}^{\rm NLP}\left(\Omega,\{\omega\};\Omega',\{\omega'\}\right)\,S_{g\bar{q}}^{\rm NLP}(\Omega',\{\omega'\};\mu_s)\notag\\
&&\hspace*{-3.5cm}
= -\left(\frac{\alpha_s}{4\pi}\right)^{\!2}\omega_1\theta(\Omega-\omega_1)\Bigg\{\bigg(4C_F\ln\frac{\Omega-\omega_1}{\omega_1}+4C_A\ln\frac{(\Omega-\omega_1)\omega_1}{\mu_s^2}
\label{eq:logmuderIC}\\
&&\hspace{-3cm}-3C_F+\beta_0\bigg)
  \delta(\omega_1-\omega_2)
-4\left(C_F-\frac{C_A}{2}\right)\,\omega_2\Gamma(\omega_1,\omega_2)\Bigg\}+\Big(\omega_1\leftrightarrow\omega_2\Big)\,.\notag
\end{eqnarray}
The logarithms of $\Omega-\omega_n$ enhance the evolution for $\omega_n$ close to $\Omega$ causing a rapid increase of the evolved soft function towards $\omega_n=\Omega$. Moreover, the soft function develops another singularity for $\omega_n = 0$ because of the logarithms in $\omega_n$. However, this singularity is suppressed compared to the one at $\omega_n = \Omega$, because the $\omega_n$ logarithms have colour factor of $C_A - C_F$, whereas the $\Omega - \omega_n$ logarithm has a colour factor $C_A + C_F$, which differs by a factor of $3$ in the large-$N_c$ limit. As a consequence, the singularity at $\omega_n=0$ is not visible for physically relevant factorization scales as shown in Fig~\ref{fig:diagonal-section}.

Another noteworthy feature of \eqref{eq:largeNcmomentumspacesol} is the $1/\Gamma(1 + 2a_{\Gamma}^+)$ term, which leads to sign oscillations in the dependence of the soft function on $\mu$, due to an overall prefactor that switches sign whenever $2a_{\Gamma}^+ \in \mathbb{Z}_{<0}$. This can be observed in the right panel of Fig.~\ref{fig:diagonal-section}, which shows how the soft function evolves for a fixed ratio of $\omega / \Omega$. The first sign change in the large-$N_c$ limit is at $\mu \approx 396.5~{\rm GeV}$, in agreement with the plot. This behaviour is mostly a leading-power effect (previously observed in LP threshold resummation of the hadronic top-pair production cross section  \cite{Beneke:2011mq}) driven by the $L_{x_0}$ dependence in the evolution kernel~\eqref{eq:NLPsoftADpos}, which determines the exponent of $x^0$ in the NLP soft function~\eqref{eq:resummed1} and \eqref{eq:VS}. The physical significance of these oscillations must be clarified in the context of the factorization formula in which the soft function appears in a convolution with jet functions and parton distributions. Beyond the first zero of the inverse gamma function, that is, for 
scale $\mu$ large enough such that $2 a_\Gamma^+<-1$, the singularity at $\omega_n=\Omega$ discussed above becomes formally non-integrable and the soft function must be amended by subtraction terms at $\omega=\Omega$ as mentioned in the previous footnote, which must also be included in the evaluation of the factorization formula. 

\subsubsection{General case}

\begin{figure}[t]
    \centering
    \begin{subfigure}[ht]{0.49\textwidth}
        \includegraphics[width=\textwidth]{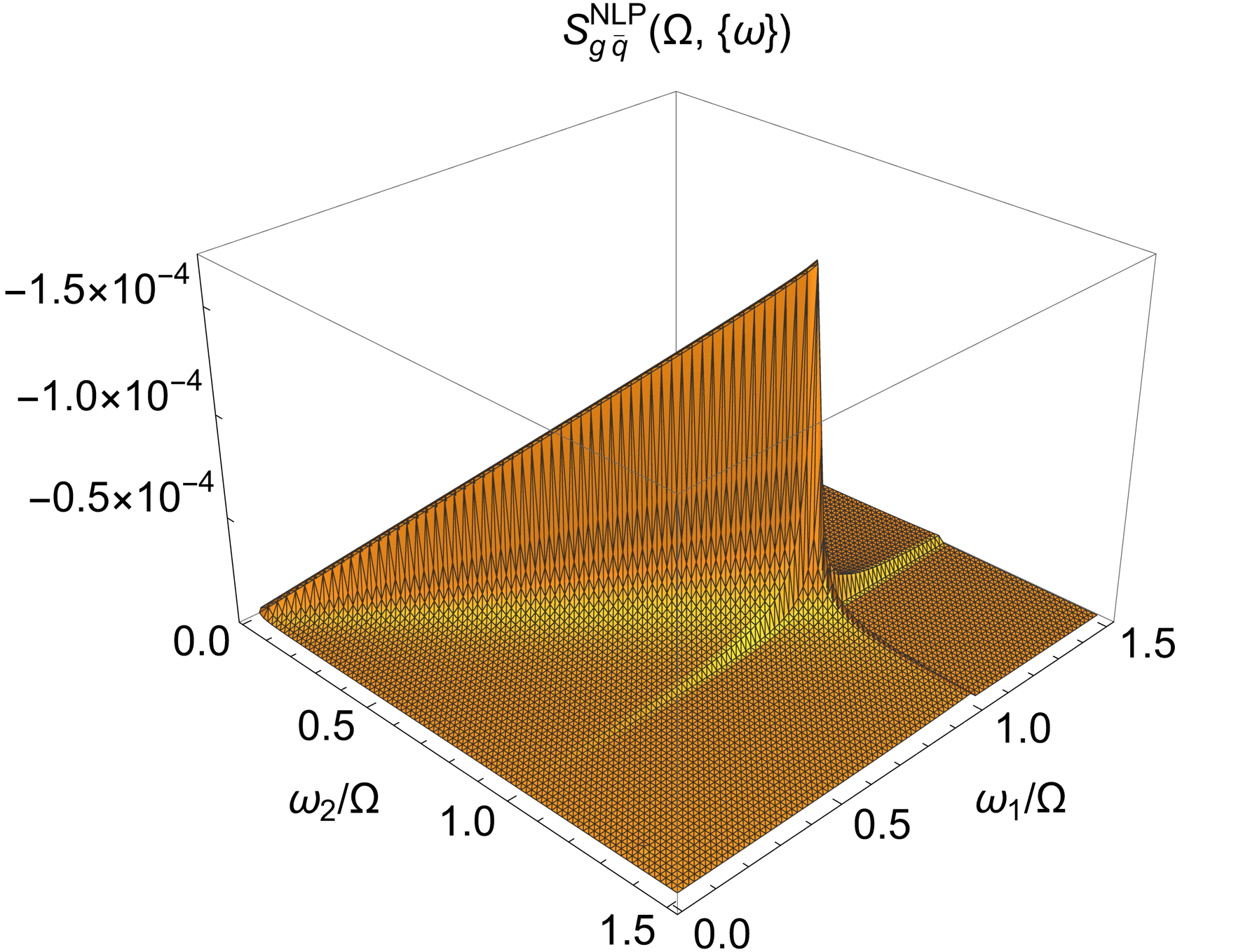}
        \label{fig:plot_3d_mu_4.1}
    \end{subfigure}
    \begin{subfigure}[ht]{0.49\textwidth}
        \includegraphics[width=\textwidth]{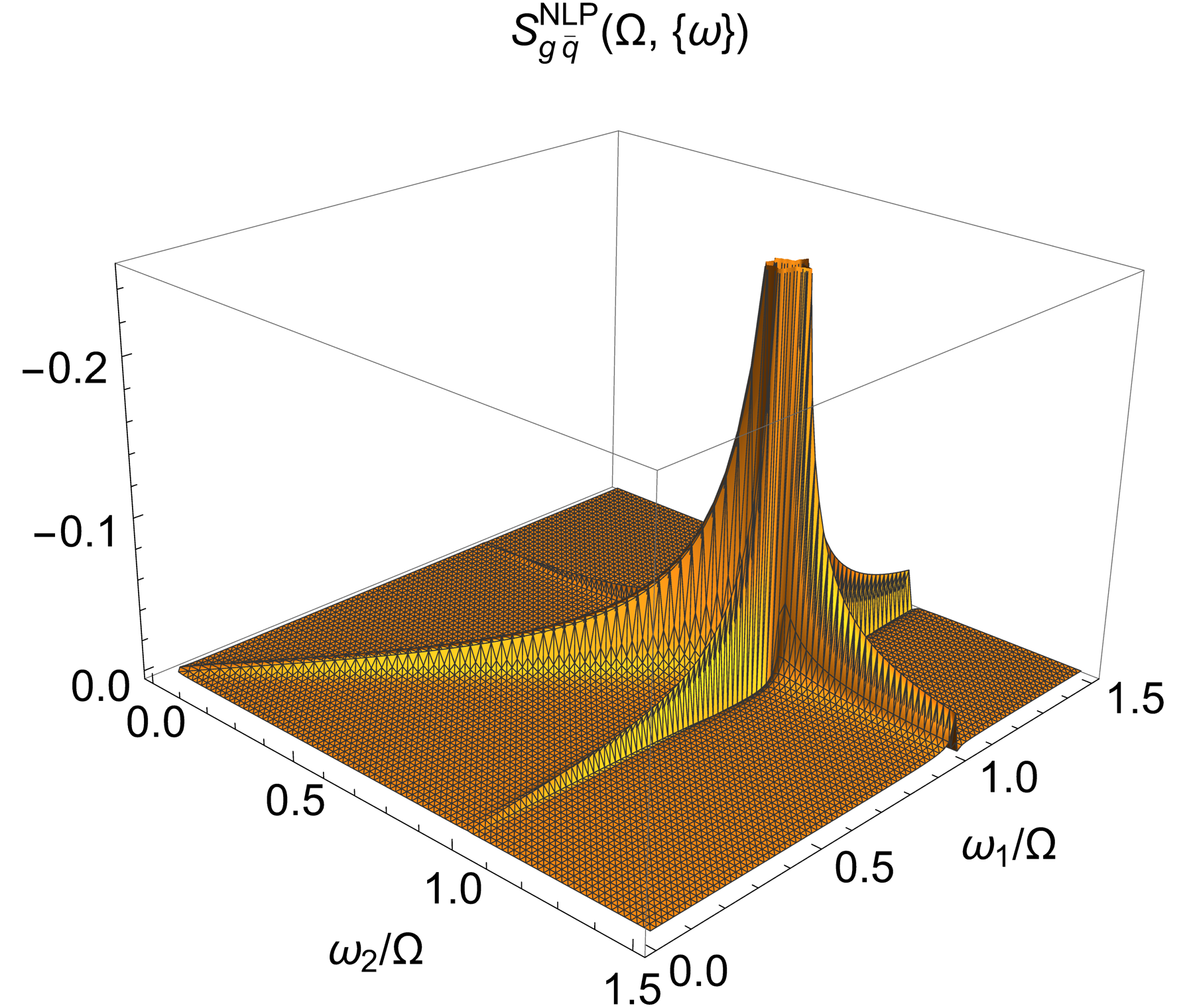}
        \label{fig:plot_3d_mu_20}
    \end{subfigure}
    \caption{3D plot of the soft function $S_{g\bar{q}}^{{\rm NLP}}(\Omega,\{\omega\})$ for initial scale $\mu_s = 10~{\rm GeV}$ and  factorization scales $\mu = 10.1~{\rm GeV}$ (left) and $\mu = 90~{\rm GeV}$ (right).}
    \label{fig:plot_3d}
\end{figure}

The singularity and oscillation features generalize to finite $N_c$. In addition, the 
evolution away from the large-$N_c$ limit shows further interesting structure of the evolved soft function, which 
is displayed for real-world QCD $N_c=3$ in Fig.~\ref{fig:plot_3d} for a value of $\mu$ 
close to the initial scale (left) and another  after the evolution to a higher scale 
(right).

%%%%%%%%%%%%%%%%%%%%%%%%%%%%%%%%%%%%%%%%%%%%%%%%%%%%%%%%%%%%%%%%%%%%%%%%%%%%%%%%%
%%%%%%%%%%%%%%%%%%%%%%%%%%%%%%%%%%%%%%%%%%%%%%%%%%%%%%%%%%%%%%%%%%%%%%%%%%%%%%%%%

At finite $N_c$, the soft function is no longer localized on the diagonal 
$\omega_1=\omega_2\leq\Omega$, but is spread out by the evolution into the entire quadrant of positive $\omega_n$. Comparing the two panels of the figure, the singular behaviour near $\omega_n=\Omega$ generated by evolution is clearly visible. In addition, the soft function 
develops a non-zero contribution for $\omega_n > \Omega$, which however is 
concentrated along the lines $\omega_1=\Omega$ and $\omega_2=\Omega$ and quickly goes to zero as $\omega_n$ increases. Technically, these non-zero contributions come from the non-vanishing value of $\rho_\Gamma=1/10$ in \eqref{eq:I-int}, which causes a series of poles or a branch cut to the right of the imaginary axis as shown in Fig.~\ref{fig:integration-contour}. Since 
$\rho_\Gamma$ is not large, the soft function still remains predominantly localized 
in the $\omega_1,\omega_2$ space around $\omega_n=\Omega$ as in the large-$N_c$ limit.

The generation of support for $\omega_n>\Omega$ is a consequence of the 
non-local distribution $\Gamma(\omega_n,\omega_n^\prime)$ in the 
anomalous-dimension kernel \eqref{eq:ADmom}. This distribution coincides with the one that appears in the renormalization of the leading-twist B-meson light-cone distribution amplitude \cite{Lange:2003ff} (technically also a soft function), which is known to generate a radiative tail for arbitrarily large values of the soft momentum fraction $\omega$. 
The concentration of the radiatively generated support on the lines  $\omega_1=\Omega$ and $\omega_2=\Omega$ can be understood from \eqref{eq:ADmom} or \eqref{eq:logmuderIC}. Referring to the latter, due to the $\Gamma(\omega_1,\omega_2)$ term in the curly bracket, the soft function develops support for $\omega_2 > \Omega$ but only (at leading order) for $\omega_1<\Omega$. Since 
the soft function is peaked in $\omega_1$ near $\Omega$, this effect is large only when $\omega_1\approx \Omega$, which explains the ridge in Fig.~\ref{fig:plot_3d} along 
$\omega_1/\Omega=1$. Due to the symmetry of the anomalous dimension, the same feature appears when $\omega_1$ and $\omega_2$ are interchanged.

Which values of $\omega_n$ are actually relevant, depends on the behaviour of the collinear functions in the convolution with the soft functions. As discussed in Sec.~\ref{sect:endpoint}, the product of collinear functions introduces a factor 
$1/(\omega_1\omega_2)$ (times powers of logarithms in higher orders of the strong coupling expansion). Along the diagonal $\omega=\omega_1=\omega_2$, this will turn the linear 
behaviour in $\omega$ into $1/\omega$ causing an endpoint divergence as $\omega\to 0$. 
This endpoint divergence should be eliminated by a similar rearrangement procedure as 
in \cite{Beneke:2022obx} and eventually absorbed into a modification of the parton distributions in the process of computing the hadronic DY cross section. The results above then show that after this rearrangement, the appropriately subtracted convolution integral with the soft function receives its main support from 
$\omega_1\approx \omega_2\approx \Omega$, which suggests that most of the energy of the soft radiation near the DY threshold in the off-diagonal parton channel is carried by the emitted soft (anti-)quark.

%%%%%%%%%%%%%%%%%%%%%%%%%%%%%%%%%%%%%%%%%%%%%%%%%%%%%%%%%%%%%%%%%%%%%%%%%%%%%%

\section{Conclusions}
\label{sect:conclusions}

In this paper, we presented for the first time the computation of the anomalous-dimension kernel of a next-to-leading-power soft function for an inclusive cross section, specifically the Drell-Yan cross section in the off-diagonal channel $g+\bar{q}\to \gamma^*+X$ near threshold. The solution to the renormalization-group equation is an essential ingredient in the resummation of threshold logarithms at sub-leading power. 

The anomalous dimension is a function of three variables, corresponding to the total energy of soft radiation and two convolution variables originating from the sub-leading power soft-quark emission Lagrangian. This function turns out to have a factorized form in the three variables, combining structures familiar from the anomalous dimension of the leading-power Drell-Yan soft function and the kernel for the amplitude-level soft function that appears in the light-quark mediated 
$gg\to h$ process \cite{Beneke:2024cpq, Liu:2022ajh}. We also establish a relation of this kernel to the soft-gluon limit of the twist-3 three-particle heavy-light light-cone operator. The three variables nevertheless get entangled in the evolution of the soft function, since the soft-function initial condition does not factorize.

The ultraviolet divergences of the off-diagonal DY soft-quark functions that give rise to the anomalous dimension are much simpler than the poles of the dimensionally regulated $O(\alpha_s^2)$ fixed-order
result \cite{Broggio:2023pbu}, which contains ultraviolet and infrared poles. 
Understanding the complete pole structure is crucial for the consistency of the next-to-leading-power factorization formula and the 
refactorization of the divergent convolution integrals 
 \cite{Beneke:2019oqx}, which remains for future work.

We solved the evolution equation for the tree-level soft function analytically in the large-$N_c$ limit and numerically for the general case. After evolution the soft function is strongly peaked near $\omega_1\approx\omega_2\approx\Omega$. It also develops support for $\omega_n>\Omega$, mainly along the lines $\omega_1=\Omega$ and $\omega_2=\Omega$. 
This suggests that after factorizing the endpoint divergence in the convolution with the jet functions into the parton distributions, most of the energy of the soft radiation near the DY threshold in the off-diagonal parton channel is carried by the emitted soft (anti-)quark.
 
%%%%%%%%%%%%%%%%%%%%%%%%%
\subsection*{Acknowledgments}
M.B. and X.W wish to thank the CERN
Theoretical Physics Department and the Munich Institute for Astro-, Particle- and Bio-Physics (MIAPbP) for hospitality while part of this work was carried out.
This work has been supported by the Excellence Cluster ORIGINS funded by the DFG under Grant No.~EXC-2094-390783311.  Y.J and X.W are also partly supported by the University Development Fund of The Chinese University of Hong Kong, Shenzhen, under Grant No.~UDF01003869 and No.~UDF01003912, respectively.

%%%%%%%%%%%%%%%%%%%%%%%%%%%%%%%%%%%%%%%%%%%%%%%%%%%%%%%%%%%%%%%%%%% 

\begin{appendix}

\section{Gauge invariance of the subtracted soft functions}
\label{appendix:gauge}

The gauge-parameter dependence of the $\delta$-regulated soft functions and subtraction terms and the  gauge invariance of the final results have been investigated in the context of \cite{Beneke:2024cpq}. The conclusion 
therein applies here as well. This appendix presents some details on 
the calculation of the DY soft function in general $R_\xi$ gauge. 

We begin with the leading-power case. The virtual correction is $\xi$-dependent in  general $R_\xi$ gauge. The calculations are rather similar to those in \cite{Beneke:2024cpq}. We quote the result here:
\begin{align}
	\label{eq:LPsoftunsvxi}
		S_{q\bar{q},{\rm uns}}^{{\rm LP}, \,v,\,(\xi)}(x_0)
		= \frac{\alpha_s C_F}{\pi} \left[-\frac{1}{\eps^2}-\frac{1}{\eps}\ln\frac{\mu^2}{\left|\delta_-\delta_+\right|}+\frac{1-\xi}{2\eps}\right]+\mathcal{O}(\eps^0),
\end{align}
where only the pole part is kept. In the Feynman gauge, we set $\xi=1$ and reproduce \eqref{eq:LPsoftunsv2}. 

The computation of the real emission contribution is more complicated, since in  general $R_\xi$ gauge, one needs to take the discontinuity of the 
$l^\mu l^\nu/l^4$ part of the propagator.  In this case, we find it simpler 
to take the discontinuity after performing the loop integration, and simplify 
the computation with the following trick: We introduce the auxiliary parameter $\Delta$ to represent the additional term in $R_\xi$ gauge of the real emission diagram $(b)$ in Fig.~\ref{fig:LP} as
\begin{eqnarray}
    \label{eq:LPsoftunsbxi}
    S_{q\bar{q},{\rm uns}}^{{\rm LP}, \,(b),\,(\xi)}(x_0)
		&=& -2\pi(1-\xi)g_s^2C_F\int\limits_{-\infty}^0\df\lambda\int\limits_{-\infty}^0\df\lambda'e^{-i\lambda\delta_+-i\lambda'\delta_-}\frac{\partial}{\partial\lambda}\frac{\partial}{\partial\lambda'}\frac{\partial}{\partial\Delta}\int\frac{\df^D l}{(2\pi)^D}\nonumber \\
 && \times \,e^{-il\cdot(x_0+\lambda n_+-\lambda'n_-)}\,\delta(l^2-\Delta)\theta(l^0)\Big|_{\Delta=0}+\mathcal{O}(\eps^0),
\end{eqnarray}
where $\partial/\partial\Delta\,\delta(l^2-\Delta)$ generates the cut propagator induced by the $(1-\xi)$-part of the covariant gluon propagator. Then it follows that the above contribution is $\mathcal{O}(\eps^0)$ as is the 
$\xi=1$ contribution. 

Next, we turn to the subtraction. In general $R_\xi$ gauge, diagrams $(a)$ and $(b)$ in Fig.~\ref{fig:LP-sub} are computed following the same strategy as for the virtual corrections. However, diagram $(c)$ is no longer zero as in Feynman gauge due to the extra tensor structure in the gluon propagator. In total, the pole part reads
\begin{equation}
	\label{eq:LPsubppolexi}
	S_{\sqcap}^{+,\,(\xi)}(t) = 1 + \frac{\alpha_s C_F}{\pi} \left[-\frac{1}{\eps^2}-\frac{1}{\eps}\left(\ln\frac{\mu}{|\delta_+|}+\ln\left(i\mu e^{\gamma_E}t\right)\right)+\frac{1-\xi}{4\eps}\right]+\mathcal{O}(\alpha_s^2,\eps^0)\,.
\end{equation}
The result for $S_{\sqcap}^{-,\,(\xi)}(t)$ follows by obvious substitutions. Then, it is clear that the gauge-dependent terms in the virtual correction cancel those in the subtraction terms.  

The case of the next-to-leading-power soft function is rather similar. The real emission diagrams do not contribute to gauge-dependent pole parts, and those in the virtual corrections cancel with those in the subtractions, shown in Fig.~\ref{fig:NLP-sub} and some diagrams that do not contribute in Feynman gauge but must now be considered. It is straightforward to perform the corresponding calculations, and we do not present them here.

%%%%%%%%%%%%%%%%%%%%%%%%%%%%%%%%%%%%%%%%%%%%%%%%%%%%%%%%%%%%%%%%%%%%%%%%%%%%%%%%%%

\section{Subtracted soft functions and parton distribution functions}
\label{appendix:subopandPDF}

In this appendix we demonstrate that the IR subtraction of the soft function that renders its anomalous dimension well-defined, when multiplied back to the collinear 
functions yields a consistent factorization formula for the Drell-Yan cross section. The essential arguments are similar to those in Sec.~4 of \cite{Beneke:2024cpq} 
for the $gg\to h$ form factor.\footnote{See also 
\cite{Chay:2013zya,Chay:2017bmy} for earlier work on this problem, although 
the details are different from the construction presented here.} 

Since 
the subtraction involves the same Wilson-line operators for the LP and 
NLP DY soft functions, it suffices to discuss the leading-power case. 
The partonic cross section for $q+\bar{q}\to \gamma^* +X$ as $z\to 1$ 
is usually written as
\begin{equation}
\label{eq:factlp}
\Delta^{\rm LP}_{q\bar{q}}(z)= H(Q^2)\,S^{\rm LP}_{q\bar{q}}(Q(1-z))
\otimes [f_{q/q}(x_1)]_{x_1\to 1}\,
[f_{\bar{q}/\bar{q}}(x_2)]_{x_2\to 1}\,.
\end{equation}
Here $S^{\rm LP}_{q\bar{q}}(\Omega)$ refers to the unsubtracted momentum-space
soft function \eqref{eq:LPsoftdef} in the present work (before introducing $\delta$ regulators). Before subtraction of the poles in 
dimensional regularization, this soft function has only IR poles, because the virtual diagrams are scaleless. There 
is no collinear radiation associated with light-cone momentum of order $Q$ into the final state $X$ for the partonic 
cross section due to the phase-space restriction near threshold.  
Hence,  there are no collinear functions, because all virtual collinear loops are scaleless as well. The (IR) poles of $H\cdot 
S^{\rm LP}_{q\bar{q}}$ in  \eqref{eq:factlp} match (twice) the 
DGLAP kernel of the quark parton distribution function in 
the $x\to 1$ limit. When computed on-shell, the loop corrections to 
the partonic parton distributions are scaleless, and $f_{q/q}(x)$ 
is given by the operator counterterm only. 
Thus \eqref{eq:factlp} yields 
the finite, renormalized partonic DY cross section near threshold.

While the content of \eqref{eq:factlp} is correct, the 
presented expression and argument are not entirely satisfactory, 
since the parton distributions are defined as matrix elements 
of collinear fields including collinear radiation into the 
final state, which is absent near threshold. 
For the present discussion, it is important to start instead from 
the factorized expression 
\begin{equation}
	\label{eq:factlpwithZchi}
	\Delta^{\rm LP}_{q\bar{q}}(z)= H(Q^2)\,S^{\rm LP}_{q\bar{q},\rm uns}(Q(1-z))\,\left|Z_\chi(p_q^2)\right|\left|Z_\chi(p_{\bar{q}}^2)\right|\,
\end{equation}
for the IR-divergent partonic cross section, 
where $Z_\chi$ is defined as 
\begin{equation}
   \langle 0|\chi^{(0)}_c(0)|q(p)\rangle = \sqrt{Z_\chi}(p^2) u_{n_+},\,\quad \langle q(p)|\bar{\chi}^{(0)}_c(0)|0\rangle = \sqrt{Z^*_\chi}(p^2) \bar{u}_{n_+}\,.
\end{equation}
Here $\chi^{(0)} = [W_c^\dagger\xi_c]^{(0)}$ denotes the collinear-gauge-invariant collinear quark field in soft-collinear effective theory (SCET) after the soft-gluon decoupling, and $u_{n_+}$ the corresponding spinor polarization. 
Alternatively, from the QCD perspective, $Z_\chi$ denotes the matrix element as defined above of the QCD quark field in light-cone gauge. The factor $Z_\chi$ accounts for the purely virtual collinear loops on the external legs 
of the DY scattering process. The above-mentioned absence of 
collinear functions is a consequence of $Z_\chi\equiv 1$ 
to all orders in perturbation theory, when the quark state is on-shell, $p^2=0$.

%%%%%%%%%%%%%%%%%%%%%%%%%%%%%%%%%%%%%%%%%%%%%%%%%%%%%%%%%%%%%%%%%%%%%%%%%%%%%%%%%
\begin{figure}[t]
\centering
\includegraphics[width=10cm]{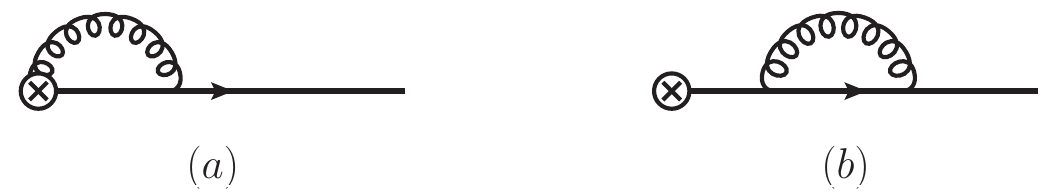}
\caption{\label{fig:Zchi}
One-loop corrections to $\langle 0 | \chi_{c}(0)|q(p)\rangle$ are given by $(a)+(b)/2$. The crossed dot denotes the $\chi$ field including the collinear 
Wilson line.}
\end{figure}
%%%%%%%%%%%%%%%%%%%%%%%%%%%%%%%%%%%%%%%%%%%%%%%%%%%%%%%%%%%%%%%%%%%%%%%%%%%%%%%%

In the following we regularize the IR divergence by a small off-shellness of the colliding partons, as already indicated in \eqref{eq:factlpwithZchi}, and focus on the UV renormalization and anomalous dimension of the soft and collinear factors in this equation. The off-shellness of the external partons propagates into the internal lines, and implies the delta regularization of the soft Wilson lines with $\delta_\pm$ determined by 
\begin{equation}
\label{eq:del2prelation}
p_q^2 = \np\cdot p_q \,\delta_-, \qquad 
p_{\bar q}^2 = \nm \cdot p_{\bar q} \,\delta_+\,.
\end{equation}
The dimensionful IR regulators \eqref{eq:del2prelation} 
introduce collinear modes with 
light-cone momentum of order $Q$ and soft-collinear modes with light-cone momentum fraction $Q(1-z)$, both of which have virtuality of order of the regulator-$p^2$.  
The former must be purely virtual, while the soft-collinear modes can be 
radiated into the final state even in threshold kinematics.
This has two important consequences. 

First, the virtual collinear loops encapsulated in $Z_\chi(p^2)$ are no longer 
scaleless. The computation of the one-loop diagrams shown in Fig.~\ref{fig:Zchi}
results in 
\begin{equation}
\label{eq:Zchi}
 \sqrt{Z_{\chi}}(p^2) = 1 + \frac{\alpha_s}{4\pi}C_F\left[\frac{2}{\eps^2}+\frac{3-4L_p}{2\eps}\right] + \mathcal{O}(\alpha_s^2,\eps^0)\,,   
\end{equation}
where $L_p= \ln\big((-p^2-i0^+)/\mu^2\big)$,
which agrees with the corresponding QED result 
\cite{Beneke:2020vnb} after substituting 
$Q_\ell^2\to C_F$. Due to \eqref{eq:del2prelation}, 
$Z_\chi$ depends effectively on the $\delta$-regulators and the large collinear momentum component of the initial parton. Moreover, 
in general covariant $R_\xi$ gauge,  $Z_\chi$ is  gauge-parameter dependent. 
Accounting for \eqref{eq:del2prelation}, we obtain
\begin{equation}
\label{eq:Zchimodulus}
 Z_{\chi}^- \equiv \left|Z_{\chi}(p_q^2)\right|  = 1 + \frac{\alpha_s}{4\pi}C_F\left[\frac{4}{\eps^2}+\frac{1}{\eps}\left(3-4\ln\frac{n_+\cdot p|\delta_-|}{\mu^2}\right)\right] + \mathcal{O}(\alpha_s^2,\eps^0)\,.
\end{equation}

Second, when introducing $\delta$-regulators, the unsubtracted soft function 
has only UV poles in $\epsilon$, which are, however, IR-regulator dependent, as shown in the main text. The dependence on the regulator indicates that the unsubtracted 
soft function $ S^{\rm LP}_{q\bar{q},\rm uns}$ is a two-scale object, containing 
soft and soft-collinear modes. The IR subtraction of the soft function is 
designed to remove the regulator-dependent soft-collinear contribution and 
combines it with the other regulator-generated objects, the collinear functions $Z_\chi(p^2)$. This is the essence of the IR rearrangement\footnote{The following equation can be read in position or momentum space with respect to the arguments of the soft functions. For the sake of brevity, we use the same notation, implicitly changing multiplications in position space to convolutions in momentum space.} 
\begin{equation}
\label{eq:refact}
H\cdot S_{q\bar{q},\rm uns}^{\rm LP}\,Z_\chi^- Z_\chi^+ = 
H\cdot \frac{S_{q\bar{q},{\rm uns}}^{\rm LP}}{S_{\sqcap}^{+}S_{\sqcap}^{-}}  \left[S_{\sqcap}^{-}Z_{\chi}^-\right]\left[S_{\sqcap}^{+}Z_{\chi}^+\right]
\end{equation}
of the factorization formula \eqref{eq:factlpwithZchi}, where $Z_\chi^-$ has been defined in \eqref{eq:Zchimodulus} and  $Z_\chi^+$ is defined similarly. 

As shown in the main text, the subtracted soft function 
$S_{q\bar{q},{\rm uns}}^{\rm LP}/(S_{\sqcap}^{+}S_{\sqcap}^{-})$ 
has a well-defined anomalous dimension (i.e. the $\delta$-regulators
drop out in the UV poles in $\epsilon$). Furthermore, it is 
gauge-parameter independent. In turn for the 
collinear factor, now including the IR subtraction, we 
multiply the expressions given in \eqref{eq:LP-subm} and 
\eqref{eq:Zchi} to obtain 
\begin{eqnarray}
Z_{\chi}^- S_{\sqcap}^-(t) &=& 1 + \frac{\alpha_s C_F}{\pi}\frac{1}{\eps}
\left[-L_t-\ln\frac{\mu}{|\delta_-|}-\ln\frac{n_+\cdot p|\delta_-|}{\mu^2}+\frac{3}{4}
\right]+\mathcal{O}(\alpha_s^2,\eps^0)\nonumber\\
        & = & 1 - \frac{\alpha_s C_F}{\pi}\frac{1}{\eps}
\left[\ln \Big(i e^{\gamma_E} \np\cdot p_q (t-i0^+)\Big)-\frac{3}{4}\right]+\mathcal{O}(\alpha_s^2,\eps^0)\,.
\label{eq:ZS}
\end{eqnarray}
Thus, the collinear factor also 
has a well-defined anomalous dimension. This demonstrates that 
the IR rearrangement restores soft-collinear factorization 
in the sense that the soft and collinear functions can be 
separately renormalized and do not depend on the artificially introduced 
regula\-tor-dependent scales.

Finally, we show that the above provides a proper derivation 
of \eqref{eq:factlp}, in other words
 \begin{equation}
\label{eq:refact2}
\underbrace{ H\cdot \frac{S_{q\bar{q},{\rm uns}}^{\rm LP}}{S_{\sqcap}^{+}S_{\sqcap}^{-}}   \left[S_{\sqcap}^{-}Z_{\chi}\right] \left[S_{\sqcap}^{+}Z_{\chi}\right]}_{\text{\eqref{eq:refact}}}{}_{\Big| \rm pole}\,\stackrel{!}{=}\,\,
 \underbrace{H\cdot S_{q\bar{q}}^{\rm LP}\,[f_{q/q}]_{x_1\to 1}\,
[f_{\bar{q}/\bar{q}}]_{x_2\to 1}}_{\text{\eqref{eq:factlp}}}{}_{\Big| \rm pole}\,,
\end{equation}
where ``pole'' refers to the pole parts in $\eps$ of the 
three separate factors. For the soft functions on both sides, this follows from 
\eqref{eq:subdef}, and since $H$ has not been modified, 
it remains to demonstrate that the Fourier transform of the pole 
part of the 
position-space result \eqref{eq:ZS} for $S_{\sqcap}^{-}Z_{\chi}$ 
coincides with the $x\to 1$ limit of the DGLAP kernel  for the 
quark parton distribution. Indeed, using \eqref{eq:FourierOmega}, we have  
(assuming $\np\cdot p_q>0$)
\begin{eqnarray}
&& - \frac{\alpha_s C_F}{\pi}\frac{1}{\eps}\,\np\cdot p_q 
\int \frac{{\rm d} t}{2\pi}\,e^{i t (1-x)\np\cdot p_q} \,\left[\ln \Big(i e^{\gamma_E} \np\cdot p_q (t-i0^+)\Big)-\frac{3}{4}\right]\nonumber\\
&&=\,  \frac{\alpha_s C_F}{2 \pi}\frac{1}{\eps} 
\left[\frac{2}{[1-x]_+} +\frac{3}{2}\delta(1-x)\right] 
= [f_{q/q}]_{x\to 1, \rm pole}\,.
\end{eqnarray}

This relation is not accidental, but a consequence 
of the factorization \cite{Korchemsky:1992xv} of the parton 
distribution function in the $x\to 1$ limit. Applied to the partonic PDF, 
the factorization formula reads 
\begin{eqnarray}
        f_{q/q}(x)\big|_{x\to 1} &=& \int\frac{\df t}{4\pi}\,e^{-i t x n_+\cdot p}\,\mbox{tr}\,\langle q(p)| \bar{\chi}_{c}(tn_+)\frac{\slashed{n}_+}{2}\, \overbrace{Y^\dagger_{n_-}(tn_+)\, [tn_+, 0]\, Y_{n_-}(0)}^{W_{\sqcap}^{-}(t)}\, \chi_{c}(0) |q(p)\rangle\nonumber\\[0.1cm]
        &=& n_+\cdot p\int\frac{\df t}{2\pi}\,e^{-i t (1-x) n_+\cdot p}\, |Z_\chi(p^2)|\, S_{\sqcap}^{-}(t)\,,
    \label{eq:partonicqpdfdef}
\end{eqnarray}
which provides another demonstration of \eqref{eq:refact2}.

%%%%%%%%%%%%%%%%%%%%%%%%%%%%%%%%%%%%%%%%%%%%%%%%%%%%%%%%%%%%%%%%%%%%%%

\section{Evolution kernel of twist-3 \texorpdfstring{$B$}{}-LCDA and NLP soft functions}
\label{app:Bt3-to-hgg}

In this appendix we establish a relation between the NLP soft-quark function anomalous dimension $\gamma_{g\bar{q}}^{\rm NLP}$, $\gamma_g$ in DY production 
and $gg\to h$ \cite{Beneke:2024cpq}, respectively, and the soft-gluon limit of the 
anomalous-dimension kernel of the twist-3 three-particle light-heavy operator~\cite{Braun:2017liq} (its matrix element defines $B$-LCDA $\phi_3$),
\begin{align}
\mathcal{O}_3(s_1,s_2) &= \bar q(s_1n_-) [s_1n_-, s_2 n_-] g_s G_{\mu\nu}(s_2n_-)[s_2n_-, 0]n_-^\nu \slashed{n}_-\gamma_\perp^\mu\gamma_5 h_v(0) \,.
\label{eq:O3}
\end{align}

To this end, we follow \cite{Beneke:2024cpq} and first express the anomalous dimension 
of the DY NLP soft-quark function in terms of the $\nm$-collinear conformal 
generators\cite{Braun:2003rp}
\begin{equation}
	\label{eq:generator}
	\widehat{S}^{(+)}_{i}=s_i^{2} \partial_{s_i}+2j s_i
	=s_i \theta_{i}+2j s_i, 
	\quad \widehat{S}^{(0)}_{i} = %
	s_i\partial_{s_i}+j
	=\theta_{i}+j,
	\quad \widehat{S}^{(-)}_{i}=-\partial_{s_i}\, ,
\end{equation}
where $j=1$ is the canonical conformal spin of the soft quark, $\widehat{S}_i$ represents the generators associated with the $s_i$ 
light-cone variable, and 
$\theta_{i}\equiv s_i\partial_{s_i}$ is the Euler operator.\footnote{These operators are the projection of the $d$-dimensional conformal-group generators onto the light-like $n_-$-direction.  The commutation relations read
\begin{equation}
	\label{eq:commutations}
	\left[\widehat{S}^{(+)}_{i}, \widehat{S}^{(-)}_{i}\right]=2 \widehat{S}^{(0)}_{i}, \quad\left[\widehat{S}^{(0)}_{i}, \widehat{S}_{i}^{(\pm)}\right]= \pm \widehat{S}_{i}^{(\pm)}\,,
\end{equation}
while $\left[\widehat{S}^{(i)}_{1}, \widehat{S}^{(j)}_{2}\right] = 0$ for $i, j\in\{0, \pm\}$.} We then find 
\begin{eqnarray}
   \label{eq:ConfNLPAD}
   \gamma_{g\bar{q}}^{\rm NLP} &=& \frac{\alpha_s}{4\pi}\Bigg[4\left(C_F-\frac{C_A}{2}\right)\left(\ln\left(\mu^2 e^{4\gamma_E}\widehat{S}^{(+)}_1\widehat{S}^{(+)}_2\right)-2\right)+2C_A\left(\ln\frac{\widehat{S}^{(-)}_1\widehat{S}^{(-)}_2}{\mu^2}-2\right)+2C_F\nonumber\\
   &&+\,2\beta_0-8(C_F+C_A)\ln\left(i\mu e^{\gamma_E}x^0/2\right) \Bigg]+\mathcal{O}(\alpha_s^2).
\end{eqnarray}
The dependence on $x^0$  cannot be expressed in 
terms of collinear conformal generators because it is not associated with any fundamental QCD field and corresponds to a total translation. The $s_i$-parts are essentially the same as those in \cite{Beneke:2024cpq} due to the common building block \eqref{eq:octetblock} of the 
corresponding soft operators.

The evolution kernel for the twist-3 three-particle light-heavy operator~\eqref{eq:O3} 
in the conformal representation takes the form~\cite{Braun:2015pha}
\begin{eqnarray}
  \label{eq:Kt3}
    \mathcal{K}_{~3} &=& \frac{2\alpha_s}{4\pi}\bigg\{2\left(C_F-\frac{C_A}{2}\right)\left[\ln\left(i\mu e^{2\gamma_E} S^{(+)}_{1}\right)
    -\frac54\right] + C_A\left[\ln\left(i \mu e^{2\gamma_E} S^{(+)}_{2}\right)-\frac12\right]\\[0.1cm]
    &&\hspace*{-1cm}+\,C_A\left[\psi\left(J_{12}+\frac32\right)+\psi\left(J_{12}-\frac32\right)-2\psi(1)-\frac34\right] -4\left(C_F-\frac{C_A}{2}\right)(-1)^{J_{12}-3/2}\frac{\Gamma\left(J_{12}-\frac32\right)} {\Gamma\left(J_{12}+\frac32\right)}\bigg\}\, ,\nonumber
\end{eqnarray}
where $\psi(x)$ is the digamma function. 
 The conformal spin operator $J_{12}$ is related to the ${\rm SL}(2,\mathbb{R})$
Casimir operator $\vec{S}_{12}^2$ acting on the two light QCD fields $\bar q(s_1n_-)$ and $G_{\mu\nu}(s_2n_-)$ in ${\cal O}_3$ by $\vec S_{12}^2 \equiv J_{12}(J_{12}-1)$, where
\begin{align}
\label{eq:S12-casimir}
   \vec S_{12}^2 = (\vec S_1 + \vec S_2)^2 \equiv \frac12\left[S_{12}^{(+)} S_{12}^{(-)} + S_{12}^{(-)} S_{12}^{(+)}\right] + S_{12}^{(0)} S_{12}^{(0)} = S_{12}^{(+)} S_{12}^{(-)} + S_{12}^{(0)} \left(S_{12}^{(0)}-1\right)
\end{align}
with $S_{12}^{(i)} = S_{1}^{(i)} + S_{2}^{(i)}$ for $i\in\{ 0, \pm\}$. 
Since $\mathcal{K}_{~3}$ 
above is computed in the NDR scheme $\{\gamma_5, \gamma^\mu\}=0$, the scale dependence of the vector and axial-vector operators coincides to all orders, hence the kernel 
${\cal K}_{~3}$ in~\eqref{eq:Kt3} also applies to \eqref{eq:O3} without 
the $\gamma_5$ matrix in the definition.  

Next, we apply the following limiting procedure: We let $s_2\mapsto \lambda s_2$, which allows us to push the location of the gluon field-strength tensor $G_{\mu\nu}(\lambda s_2n_-)$ to $-\infty n_-$ by taking  $\lambda\to-\infty$ (assuming  $s_2>0$ without loss of generality). The two Wilson lines connecting to $G_{\mu\nu}$ are maintained and extended to $-\infty n_-$. This decouples the gluon field from the operator ${\cal O}_3$ as a consequence of the $i0^+$ prescriptions \eqref{eq:delta} for the infinite Wilson lines, which exponentially suppress contributions at infinity. The \textit{reduced} operator therefore factorizes,
\begin{align}\label{eq:O3-red}
 \mathcal{O}^{\rm red}_3 &= \left\{\bar q(s_1n_-) [s_1n_-, -\infty n_-]T^a\slashed{n}_- [-\infty n_-,0] h_v(0)\right\} \left(n_-^\nu\gamma_\perp^\mu g_s G^a_{\mu\nu}(-\infty n_-)\right)\notag\\
 &\equiv \mathcal{O}_h(s_1)\mathcal{O}_{g}\,.
\end{align}
The decoupling of $G^a_{\mu\nu}$ implies that $\mathcal{O}_{g}\equiv n_-^\nu\gamma_\perp^\mu g_s G^a_{\mu\nu}(-\infty n_-)$ can only generate a (potentially infinite) constant $\gamma_G$ 
in the evolution kernel from the self-energy of the field-strength tensor,  the scale dependence of the coupling $g_s$, and the gluon-Wilson line interaction.  Note that both operators $\mathcal{O}_h(s_1)$ and $\mathcal{O}_{g}$ are formally gauge invariant, since the gauge transformation at infinity is trivial, yet the separation into the two factors renders the separate evolution kernels for the two resulting operators gauge-dependent (because $\gamma_G$ is gauge-dependent).  
This is expected as operators with infinite Wilson lines regulated by the $\delta_\pm$-regulators are gauge-dependent.  Gauge invariance is only restored after dividing out the subtraction operator as explicitly demonstrated in~\cite{Beneke:2024cpq} and the present work. From the above discussion we see that the gauge dependence must be a pure constant. Therefore, focusing on the gauge-invariant non-constant part of the evolution kernel in the following allows us to drop the $O_g$ part from ${\cal O}^{\rm red}_3$. 
If we further replace $h_v(0)$, which acts like a semi-infinite time-like Wilson line, by a light-like one extending in a direction different from $n_-$, we obtain, 
choosing the direction $\np$, 
\begin{align}\label{eq:O3np}
\mathcal{O}_{h}(s_1) \to \mathcal{O}_{n_+}(s_1) &=\bar q(s_1n_-) [s_1n_-, -\infty n_-]T^a\slashed{n}_- [-\infty n_-,0] [0, -\infty n_+]\, .
\end{align}
The right-hand side is identical to the $\mathbf{\bar{T}}$ part of the NLP DY soft-quark operator~\eqref{eq:NLPsoftoperatordef1} after a total translation and the replacement $s_2\to s_1$, which do not affect the evolution kernel. We note that the evolution kernels for $\mathcal{O}_h$ in~\eqref{eq:O3-red} and $\mathcal{O}_{\np}$ differ only by a constant partially related to the heavy quark renormalization $\gamma_h$ and $\ln\delta_\pm$, hence the non-constant part of the kernel is not affected by the replacement \eqref{eq:O3np}. This leads to the main conclusion of this appendix: {\em the kernel of the  twist-3 three-particle light-heavy operator in the soft-gluon limit coincides with ``half'' of the anomalous dimensions $\gamma_{g\bar{q}}^{\rm NLP}$, $\gamma_g$ of the NLP soft-quark functions  in DY production 
and $gg\to h$ up to a constant.}

To demonstrate this explicitly for the one-loop kernel $\mathcal{K}_3$, 
we let $s_2\mapsto \lambda s_2$, which implies (see also \cite{Braun:2018fiz})
\begin{align}
S_{2}^{(+)}  \mapsto \lambda S_{2}^{(+)}  \, ,\qquad
S_{2}^{(0)} \mapsto S_{2}^{(0)}\, ,\qquad
S_{2}^{(-)} \mapsto \lambda^{-1} S_{2}^{(-)} \, ,
\end{align}
for the conformal generators, and yields
\begin{align}
\label{eq:S12-limit}
S_{12}^{(+)} \mapsto \lambda S_{2}^{(+)} +{\cal O}(1)\, , \qquad  
S_{12}^{(0)} = S_{1}^{(0)} + S_{2}^{(0)}\, ,\qquad
S_{12}^{(-)} \mapsto S_{1}^{(-)} + {\cal O}(\lambda^{-1})
\end{align}
 as $\lambda\to-\infty$. Using $\vec S_{12}^2 \equiv J_{12}(J_{12}-1)$, 
\eqref{eq:S12-casimir} and~\eqref{eq:S12-limit}, we find
\begin{align}
&\psi\left(J_{12}+\frac32\right)+\psi\left(J_{12}-\frac32\right) \stackrel{\lambda\to-\infty} \mapsto \ln\left(\lambda S_1^{(-)}S_2^{(+)}\right) + O(\lambda^{-1})\notag\\
&\hspace*{2cm}\mapsto \ln\left(\mu^{-1} S_1^{(-)} \right) + \ln(\mu\lambda S_2^{(+)}) + O(\lambda^{-1})\,,\\[0.1cm]
&(-1)^{J_{12}-3/2}\frac{\Gamma\left(J_{12}-\frac32\right)} {\Gamma\left(J_{12}  +\frac32\right)} \mapsto O(\lambda^{-1})\, ,
\end{align}
where the $\overline{\rm MS}$-scale $\mu$ is introduced for dimensional reasons when separating the dependence on $S_1^{(-)}$, $S_2^{(+)}$. 
is results in the  reduced kernel\footnote{Here constants are again dropped. Also note that $\lambda S_2^{(+)}$ must be considered as a whole, hence the factor of $\lambda$ in 
$\ln\left(\lambda S_2^{(+)} \right)$ cannot be dropped as a constant term.  }
\begin{align}\label{eq:t3-red}
{\cal K}_{~3}^{\rm red} &=\frac{2\alpha_s}{4\pi}\left\{ 2\left(C_F-\frac{C_A}{2}\right)\ln\left(i \mu e^{2\gamma_E} S^{(+)}_{1}\right) +C_A \ln\left(\frac{S^{(-)}_{1}}{\mu}\right)  
+2C_A\ln\left(\mu\lambda S_2^{(+)}\right)\right\}\, .
\end{align} 
The additive dependence on $S^{(\pm)}_{1}$, $S^{(+)}_{2}$ indicates that the gluon and light-quark contribution to the evolution kernel of ${\cal O}_3^{\rm red}$ decouple. Since the gluon field resides at infinity, its evolution is expected to be autonomous and fully captured by $\ln(\mu\lambda S_2^{(+)})$, while the $s_1$-dependent part is caused only by the light quark.
The operator $\ln (\mu\lambda S_2^{(+)})$ is present in $\mathcal{K}_{~3}^{\rm red}$, since $\mathcal{O}_g$ in~\eqref{eq:O3-red} still interacts with the Wilson lines 
and, hence, the $\mathcal{O}_g$-part does not directly vanish. However, in the $\lambda\to-\infty$ limit this term  becomes an infinite constant (``$\ln(\infty)$'') which acts trivially on $O_3^{\rm red}$. Since we focus on the non-constant part,  the $\ln(\mu\lambda S_2^{(+)})$ term in \eqref{eq:t3-red} can be dropped. From the above general discussion, $\mathcal{K}_{~3}^{\rm red}$ with this term removed should equal {\em exactly} the $s_1$ part of the one-loop kernel for $O_{n_+}(s_1)$ in~\eqref{eq:O3np} and hence the NLP  DY soft function, i.e. the latter should be  
\begin{align}\label{eq:gamma_gq_v}
    \gamma_{g\bar{q},s_1}^{\rm NLP,{\rm virtual}} &=\frac{2\alpha_s}{4\pi}\left\{ 2\left(C_F-\frac{C_A}{2}\right)\ln\left(i \mu e^{2\gamma_E} S^{(+)}_{1}\right) +C_A \ln\left(\frac{S^{(-)}_{1}}{\mu}\right)  \right\} +{\rm constant}\,.
\end{align}
Compared to~\eqref{eq:ConfNLPAD}, we see that this is indeed the case. The other $s_2$-dependent part can be obtained from $S^{(\pm)}_1 \to S^{(\pm)}_2$.  
The constant term and $x^0$-dependent parts cannot be determined in this way, but can, for example, be inferred from  RG-consistency of the factorization formula for a physical process and the universality of the cusp anomalous dimensions induced by connecting Wilson lines of different directions.

%%%%%%%%%%%%%%%%%%%%%%%%%%%%%%%%%%%%%%%%%%%%%%%%%%%%%%%%%%%%%%%%%%%%%%

\section{Renormalization-group functions}
\label{appendix:RG}

The QCD $\beta$-function is defined as
\begin{equation}
    \frac{{\rm d} \alpha_s(\mu)}{{\rm d}\ln\mu} = \beta(\alpha_s(\mu)) = -2\eps \alpha_s -2\alpha_s\sum_{n\geq 0}\left(\frac{\alpha_s}{4\pi}\right)^{n+1}\beta_n.
\end{equation}
The perturbative expansions of the two RG functions defined in \eqref{eq:RGfunc} read
\begin{align}
   \label{eq:RGfuncexpansion}
   S_\Gamma(\nu,\mu) &= \frac{\Gamma_0}{4 \beta_0^2}\left\{\frac{4 \pi}{\alpha_s(\nu)}\left(1-\frac{1}{r}-\ln r\right)+\left(\frac{\Gamma_1}{\Gamma_0}-\frac{\beta_1}{\beta_0}\right)(1-r+\ln r)+\frac{\beta_1}{2 \beta_0} \ln ^2 r+\mathcal{O}(\alpha_s)\right\},\notag\\[0.2cm]
    a_\gamma(\nu,\mu) &= \frac{\gamma_0}{2 \beta_0}\ln  r+\mathcal{O}(\alpha_s),
\end{align}
where here $r=\alpha_s(\mu) / \alpha_s(\nu)$ and 
\begin{equation}
\Gamma_{\rm cusp}(\alpha_s) = \sum_{n\geq 0}\left(\frac{\alpha_s}{4\pi}\right)^{n+1}\Gamma_n,\qquad \gamma(\alpha_s) = \sum_{n\geq 0}\left(\frac{\alpha_s}{4\pi}\right)^{n+1}\gamma_n\,.
\end{equation}
To the two-loop order $\Gamma_{\rm cusp}^R=C_R\gamma_{\rm cusp}$ satisfies Casimir scaling and the first two coefficients of $\gamma_{\rm cusp}$ are  given by \cite{Korchemskaya:1992je}
($n_f$ the number of active quark flavours)
\begin{equation}
\gamma_{\rm cusp,0} = 4\,,\qquad 
\gamma_{\rm cusp,1} = C_A\left(\frac{268}{9}-\frac{4\pi^2}{3}\right)-\frac{40n_f}{9}\,.
\end{equation}

\end{appendix}

% -----------------------------------------------------------------------------
% references
{%\footnotesize
\bibliography{biblio}
\bibliographystyle{JHEP.bst}
}

\end{document}